\documentclass[aps,twocolumn,superscriptaddress,floatfix,nofootinbib,showpacs,amsmath,amssymb,altaffilletter]{revtex4-1}
\input{./ds.def}

\begin{document}
\title{DarkSide-50 532-day Dark Matter Search with Low-Radioactivity Argon}
\date{\today}
\newcommand{\Alberta}{Department of Physics, University of Alberta, Edmonton, AB T6G 2R3, Canada}
\newcommand{\APC}{APC, Universit\'e Paris Diderot, CNRS/IN2P3, CEA/Irfu, Obs de Paris, USPC, Paris 75205, France}
\newcommand{\AQLNGS}{INFN Laboratori Nazionali del Gran Sasso, Assergi (AQ) 67100, Italy}
\newcommand{\AQGSSI}{Gran Sasso Science Institute, L'Aquila 67100, Italy}
\newcommand{\AUM}{InstitutodeF\'õsica,Universidad Nacional Auto\'nomade M\'exico(UNAM), M\'exico 01000, Mexico}
\newcommand{\Augustana}{Physics Department, Augustana University, Sioux Falls, SD 57197, USA}
\newcommand{\Belgorod}{Radiation Physics Laboratory, Belgorod National Research University, Belgorod 308007, Russia}
\newcommand{\BHSU}{School of Natural Sciences, Black Hills State University, Spearfish, SD 57799, USA}
\newcommand{\BINP}{Budker Institute of Nuclear Physics, Novosibirsk 630090, Russia}
\newcommand{\BNL}{Brookhaven National Laboratory, Upton, NY 11973, USA}
\newcommand{\BOINFN}{INFN Bologna, Bologna 40126, Italy}
\newcommand{\BOUniPHY}{Physics Department, Universit\`a degli Studi di Bologna, Bologna 40126, Italy}
\newcommand{\CAUniCHE}{Department of Mechanical, Chemical, and Materials Engineering, Universit\`a degli Studi, Cagliari 09042, Italy}
\newcommand{\CAUniPHY}{Physics Department, Universit\`a degli Studi di Cagliari, Cagliari 09042, Italy}
\newcommand{\CAINFN}{INFN Cagliari, Cagliari 09042, Italy}
\newcommand{\Carleton}{Department of Physics, Carleton University, Ottawa, ON K1S 5B6, Canada}
\newcommand{\Campinas}{Physics Institute, Universidade Estadual de Campinas, Campinas 13083, Brazil}
\newcommand{\CentroFermi}{Museo della fisica e Centro studi e Ricerche Enrico Fermi, Roma 00184, Italy}
\newcommand{\CIEMAT}{CIEMAT, Centro de Investigaciones Energ\'eticas, Medioambientales y Tecnol\'ogicas, Madrid 28040, Spain}
\newcommand{\Cluj}{National Institute for R\&D of Isotopic and Molecular Technologies, Cluj-Napoca, 400293, Romania}
\newcommand{\CTLNS}{INFN Laboratori Nazionali del Sud, Catania 95123, Italy}
\newcommand{\ENUniCEE}{Engineering and Architecture Faculty, Universit\`a di Enna Kore, Enna 94100, Italy}
\newcommand{\ETHZ}{Institute for Particle Physics, ETH Z\"urich, Z\"urich 8093, Switzerland}
\newcommand{\FNAL}{Fermi National Accelerator Laboratory, Batavia, IL 60510, USA}
\newcommand{\FortLewis}{Department of Physics and Engineering, Fort Lewis College, Durango, CO 81301, USA}
\newcommand{\GEUni}{Physics Department, Universit\`a degli Studi di Genova, Genova 16146, Italy}
\newcommand{\GEINFN}{INFN Genova, Genova 16146, Italy}
\newcommand{\GlenEllyn}{Glen Ellyn, Illinois 60137, USA}
\newcommand{\Hawaii}{Department of Physics and Astronomy, University of Hawai'i, Honolulu, HI 96822, USA}
\newcommand{\Houston}{Department of Physics, University of Houston, Houston, TX 77204, USA}
\newcommand{\IHEP}{Institute of High Energy Physics, Beijing 100049, China}
\newcommand{\IPNO}{Institut de Physique Nucl\`eaire dÕOrsay, 91406, Orsay, France}
\newcommand{\INSTM}{Interuniversity Consortium for Science and Technology of Materials, Firenze 50121, Italy}
\newcommand{\IPHC}{IPHC, Universit\'e de Strasbourg, CNRS/IN2P3, Strasbourg 67037, France}
\newcommand{\JINR}{Joint Institute for Nuclear Research, Dubna 141980, Russia}
\newcommand{\Krakow}{M. Smoluchowski Institute of Physics, Jagiellonian University, 30-348 Krakow, Poland}
\newcommand{\Kurchatov}{National Research Centre Kurchatov Institute, Moscow 123182, Russia}
\newcommand{\Laurentian}{Department of Physics and Astronomy, Laurentian University, Sudbury, ON P3E 2C6, Canada}
\newcommand{\LNFINFN}{INFN Laboratori Nazionali di Frascati, Frascati 00044, Italy}
\newcommand{\Lodz}{Institute of Applied Radiation Chemistry, Lodz University of Technology, 93-590 Lodz, Poland}
\newcommand{\LPNHE}{LPNHE, CNRS/IN2P3, Sorbonne Universit\'e, Universit\'e Paris Diderot, Paris 75252, France}
\newcommand{\Manchester}{The University of Manchester, Manchester M13 9PL, United Kingdom}
\newcommand{\MEPhI}{National Research Nuclear University MEPhI, Moscow 115409, Russia}
\newcommand{\MIBIINFN}{INFN Milano Bicocca, Milano 20126, Italy}
\newcommand{\MIINFN}{INFN Milano, Milano 20133, Italy}
\newcommand{\MIPoliICA}{Civil and Environmental Engineering Department, Politecnico di Milano, Milano 20133, Italy}
\newcommand{\MIPoliCHE}{Chemistry, Materials and Chemical Engineering Department ``G.~Natta", Politecnico di Milano, Milano 20133, Italy}
\newcommand{\MIPoliEIB}{Electronics, Information, and Bioengineering Department, Politecnico di Milano, Milano 20133, Italy}
\newcommand{\MIPoliENE}{Energy Department, Politecnico di Milano, Milano 20133, Italy}
\newcommand{\MIUni}{Physics Department, Universit\`a degli Studi di Milano, Milano 20133, Italy}
\newcommand{\MSU}{Skobeltsyn Institute of Nuclear Physics, Lomonosov Moscow State University, Moscow 119234, Russia}
\newcommand{\NAINFN}{INFN Napoli, Napoli 80126, Italy}
\newcommand{\NAUniPHY}{Physics Department, Universit\`a degli Studi ``Federico II'' di Napoli, Napoli 80126, Italy}
\newcommand{\NAUniCHE}{Chemical, Materials, and Industrial Production Engineering Department, Universit\`a degli Studi ``Federico II'' di Napoli, Napoli 80126, Italy}
\newcommand{\NSU}{Novosibirsk State University, Novosibirsk 630090, Russia}
\newcommand{\OACINAF}{INAF Osservatorio Astronomico di Capodimonte, 80131 Napoli, Italy}
\newcommand{\Petersburg}{Saint Petersburg Nuclear Physics Institute, Gatchina 188350, Russia}
\newcommand{\PGUniCBB}{Chemistry, Biology and Biotechnology Department, Universit\`a degli Studi di Perugia, Perugia 06123, Italy}
\newcommand{\PGINFN}{INFN Perugia, Perugia 06123, Italy}
\newcommand{\PIINFN}{INFN Pisa, Pisa 56127, Italy}
\newcommand{\PIUniPHY}{Physics Department, Universit\`a degli Studi di Pisa, Pisa 56127, Italy}
\newcommand{\PNNL}{Pacific Northwest National Laboratory, Richland, WA 99352, USA}
\newcommand{\Princeton}{Physics Department, Princeton University, Princeton, NJ 08544, USA}
\newcommand{\Queens}{Department of Physics, Engineering Physics and Astronomy, QueenÕs University, Kingston, ON K7L 3N6, Canada}
\newcommand{\RHUL}{Department of Physics, Royal Holloway University of London, Egham TW20 0EX, UK}
\newcommand{\RMTreINFN}{INFN Roma Tre, Roma 00146, Italy}
\newcommand{\RMTreUni}{Mathematics and Physics Department, Universit\`a degli Studi Roma Tre, Roma 00146, Italy}
\newcommand{\RMUnoINFN}{INFN Sezione di Roma, Roma 00185, Italy}
\newcommand{\RMUnoUni}{Physics Department, Sapienza Universit\`a di Roma, Roma 00185, Italy}
\newcommand{\SAINFN}{INFN Salerno, Salerno 84084, Italy}
\newcommand{\SNOLabaddress}{SNOLAB, Lively, ON P3Y 1N2, Canada}
\newcommand{\SNOLAB}{SNOLAB, Lively, ON P3Y 1N2, Canada}
\newcommand{\SSUniCHP}{Chemistry and Pharmacy Department, Universit\`a degli Studi di Sassari, Sassari 07100, Italy}
\newcommand{\Sussex}{Physics and Astronomy, University of Sussex, Brighton BN1 9QH, UK}
\newcommand{\Temple}{Physics Department, Temple University, Philadelphia, PA 19122, USA}
\newcommand{\TNFBK}{Fondazione Bruno Kessler, Povo 38123, Italy}
\newcommand{\TNTIFPA}{Trento Institute for Fundamental Physics and Applications, Povo 38123, Italy}
\newcommand{\TNUni}{Physics Department, Universit\`a degli Studi di Trento, Povo 38123, Italy}
\newcommand{\TOINFN}{INFN Torino, Torino 10125, Italy}
\newcommand{\TOPoli}{Department of Electronics and Communications, Politecnico di Torino, Torino 10129, Italy}
\newcommand{\TOUni}{Physics Department, Universit\`a degli Studi di Torino, Torino 10125, Italy}
\newcommand{\TRIUMFaddress}{TRIUMF, 4004 Wesbrook Mall, Vancouver, British Columbia V6T2A3, Canada}
\newcommand{\TUM}{Physik Department, Technische Universit\"at M\"unchen, Munich 80333, Germany}
\newcommand{\UB}{Universiatat de Barcelona, Barcelona E-08028, Catalonia, Spain} 
\newcommand{\UCDavis}{Department of Physics, University of California, Davis, CA 95616, USA}
\newcommand{\UCLA}{Physics and Astronomy Department, University of California, Los Angeles, CA 90095, USA}
\newcommand{\UMass}{Amherst Center for Fundamental Interactions and Physics Department, University of Massachusetts, Amherst, MA 01003, USA}
\newcommand{\UOC}{Department of Chemistry, University of Crete, P.O. Box 2208, 71003 Heraklion, Crete, Greece}
\newcommand{\USP}{Instituto de F\'isica, Universidade de S\~ao Paulo, S\~ao Paulo 05508-090, Brazil}
\newcommand{\VTech}{Virginia Tech, Blacksburg, VA 24061, USA}
\author{P.~Agnes}\affiliation{\Houston}
\author{I.F.M.~Albuquerque}\affiliation{\USP}
\author{T.~Alexander}\affiliation{\PNNL}
\author{A.K.~Alton}\affiliation{\Augustana}
\author{G.R.~Araujo}\affiliation{\USP}
\author{M.~Ave}\affiliation{\USP}
\author{H.O.~Back}\affiliation{\PNNL}
\author{B.~Baldin}\altaffiliation{Present address: Raleigh, NC 27613-3313, USA}\affiliation{\FNAL}
\author{G.~Batignani}\affiliation{\PIINFN}\affiliation{\PIUniPHY}
\author{K.~Biery}\affiliation{\FNAL}
% ...

\collaboration{The \DS\ Collaboration}\noaffiliation
\begin{abstract}
The \DSf\ direct-detection dark matter experiment is a dual-phase argon time projection chamber operating at Laboratori Nazionali del Gran Sasso.  This paper reports on the blind analysis of a \DSfDdExposure\  exposure using a target of low-radioactivity argon extracted from underground sources.  We find no events in the dark matter selection box and set a \SI{90}{\percent} C.L. upper limit on the dark matter-nucleon spin-independent cross section of \DSfDdHundredGeVLim\ (\DSfDdOneTeVLim, \DSfDdTenTeVLim)
for a WIMP mass of \SI{100}{\GeV\per\square\c} (\SI{1}{\TeV\per\square\c}, \SI{10}{\TeV\per\square\c}).
\end{abstract}
\pacs{29.40.Gx, 95.35.+d, 95.30.Cq,  95.55.Vj}
\keywords{Dark matter, \WIMPs, Noble liquid detectors, Low-background detectors, Liquid scintillators, Blind analysis}
\maketitle
Despite much evidence from astronomy for dark matter (\DM), years of laboratory and indirect searches have yielded no experimental evidence for DM that is not contradicted by other experiments.  Weakly Interacting Massive Particles (\WIMPs) remain a promising candidate for DM, but direct searches are being pushed to probe lower \WIMP-nuclear interaction cross sections and to lower (\SI{<10}{\GeV\per\square\c}) and higher (\SI{>1}{\TeV\per\square\c}) \DM\ masses.  Probing lower cross sections requires higher sensitivity and hence larger exposures (target mass$\times$run time), and also, as importantly, more efficient background discrimination.  This issue is especially acute for spin-independent scattering for \DM\ masses above \SI{10}{\GeV\per\square\c}, where current limits on the \WIMP-nucleon cross section are \SI{<E-44}{\square\cm}, reaching as low as \SI{4.1E-47}{\square\cm} at \SI{30}{\GeV\per\square\c}~\cite{Aprile:2018vb}.

Liquid argon Time Projection Chambers (\LArTPCs) share the scalability and 3D position reconstruction of liquid xenon \TPCs. Moreover, \LArTPCs\ have powerful pulse shape discrimination (\PSD) in the scintillation channel that separates the nuclear recoils (\NR) expected from \WIMP\ scattering from the electron recoil (\ER) events from the dominant $\beta$- and $\gamma$-induced backgrounds.  Exploiting this \PSD, the single-phase \DEAP\ \LAr\ scintillation detector has recently reported the best available \DM-nucleon cross section limit using an \ce{Ar} target, \SI{1.2E-44}{\square\cm} at a \DM\ mass of \SI{100}{\GeV\per\square\c}, from an initial 9.87 tonne-day exposure~\cite{Amaudruz:2018gr}.

In this paper, we report results from a \DSfDdLTPostQualCutnum-live-day exposure of \DSf, a \LArTPC\ with an active mass of \DSfActiveMass\ of low-radioactivity argon from underground sources (\UAr) deployed in a liquid-scintillator veto (\LSV) for neutron and $\gamma$-ray rejection and a water Cherenkov veto (\WCV) for shielding and muon detection.  We report here the most sensitive result to date with an argon target and demonstrate the effectiveness of this combination of detectors in rejecting a broad range of backgrounds.  This paper describes the techniques developed for a blind analysis of the \DSfDdLTPostQualCutnum-live-day data set, which required detailed prediction of the background and deployment of new rejection methods.

\section{The \DSf\ Detectors}
\label{sec:Detectors}

The \DSf\ experiment is located in Hall~C of the Gran Sasso National Laboratory (\LNGS) in Italy, at a depth of \LNGSDepthMWE~\cite{Bellini:2013kr}.

\begin{figure}[t!]
\centering
\includegraphics[width=0.5\textwidth]{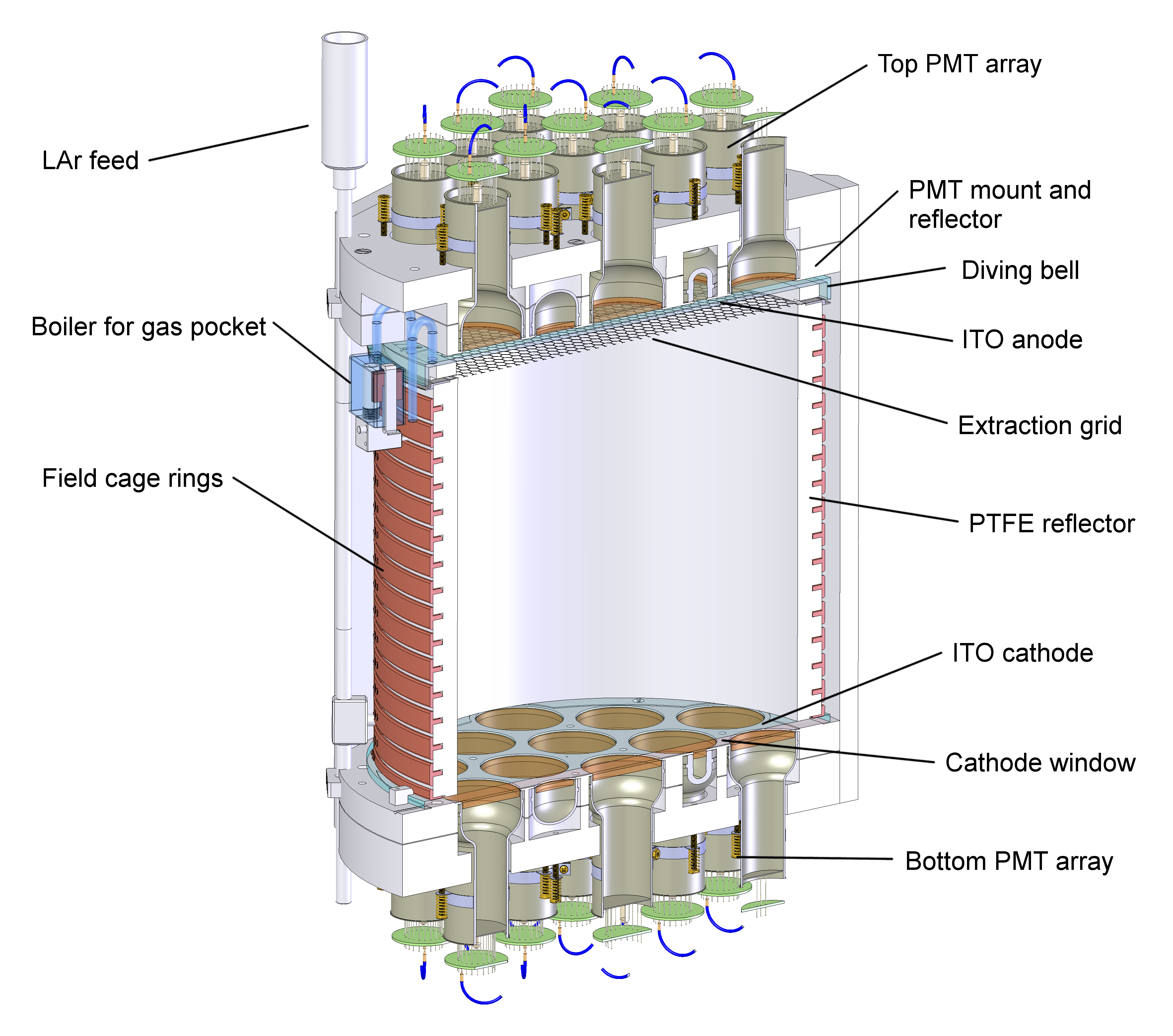}
\caption{The \DSf\ TPC.  Reproduced from \cite{Agnes:2015gu} under \cite{CCBY}.
} 
\label{fig:DS50TPC}
\end{figure}

The \DSf\ \DM\ detector is a two-phase (liquid and gas) argon TPC, described  in Ref.~\cite{Agnes:2015gu} and shown schematically in Fig.~\ref{fig:DS50TPC}.  Briefly, a cylindrical volume containing \UAr\ is viewed through fused-silica windows by top and bottom arrays of \DSfPMTsNumberTop\ \DSfPMTSize\ \DSfPMT\ photomultiplier tubes (\PMTs).  The windows are coated with Indium-Tin-Oxide (ITO) which acts as the cathode (bottom) and anode (top) of the \TPC.  The \PMTs\ operate immersed in \LAr\ and are fitted with cryogenic preamplifiers~\cite{Agnes:2017ck}.  The pre-amplifiers allow operation at reduced PMT gain, taming breakdown issues in these PMTs.

\LAr\ is boiled to form a \DSfGArHeight-thick gas pocket under the anode window.  A grid \DSfLArAboveMeshHeight\ beneath the liquid-gas interface separates a \DSfDriftField\ drift region in the main active volume from a higher-field extraction region.  

The side wall of the active LAr volume is a Teflon reflector.  The inner surfaces of the Teflon reflector and the windows are coated with tetraphenylbutadiene (\TPB), which shifts the \ArWaveLength\ argon scintillation light to \TPBWaveLength, allowing transmission through the windows and detection by the \PMTs.

Interactions in the active volume result in ER or NR events which produce primary scintillation (\SOne) as well as ionization in the LAr.  Ionization electrons surviving recombination at the event site are drifted to the liquid-gas interface, where the extraction field injects them into the gas region.  In the gas, the electric field is large enough to cause the electrons to produce a second signal (\STwo) by gas proportional scintillation.  \SOne\ and \STwo\ are both measured with the PMT arrays.  \SOne\ (or, for higher resolution, a linear combination of \SOne\ and \STwo) measures energy; the drift time ($t_{\rm drift}$), the time between the detection of \SOne\ and \STwo, measures the vertical ($z$) location of the event; and the pattern of \STwo\ on the \PMT\ arrays measures the $x$ and $y$ coordinates of the event.

The \DSf\ veto system is described in detail in Ref.~\cite{Agnes:2016fw}.  The \LSV\ is filled with \LSVScintillatorMass\ of borated liquid scintillator that detects neutrons via both prompt signals from thermalization and delayed signals from capture products.  It detects neutrons producing NR in the \LArTPC\ with extremely high efficiency (see Sec.~\ref{sec:Radiogenic}) and also detects about a third of the $\gamma$-rays giving \ER\ in the \TPC.  The \LSV\ is surrounded by the \CTFWaterMass\ \WCV, which provides shielding for the \LSV\ and a veto for cosmic ray muons.  Radioactive calibration sources for the characterization of the \TPC\ and \LSV\ are deployed through the \WCV\ and \LSV\ to the side of the cryostat using an articulated arm described in Ref.~\cite{Agnes:2017ec}. 

Under normal running conditions for the \WIMP\ search, all three detectors are read out upon a trigger from the \TPC\ that requires at least two PMTs above a threshold of \DSfTriggerDiscriminatorThreshold~\cite{Agnes:2017ck}.  Much of the data reported here also includes \DSfPulserTriggerRate\ of pulser-generated triggers, which provides an unbiased sample of detector baselines and signals.
Timestamps are recorded with the data from each detector to allow later synchronization.

\section{Data Description and Calibration}
\label{sec:DataDescription}

Data are recorded from the TPC and both vetoes with each trigger.  \TPC\ data contains the waveforms from the \DSfPMTsNumber\ \PMTs, digitized at \DSfADCSamplingRate\ without zero suppression~\cite{Agnes:2017ck}.  The digitized waveforms are acquired in a single \DSfTriggerAcquisitionTotalWindow\ window, beginning \DSfTriggerAcquisitionPreWindow\  before the trigger time and long enough to include \SOne\ and \STwo, given the maximum electron drift time of \DSfDriftTimeMax.  Data from each \PMT\ in both vetoes are digitized at \ODADCSamplingRate\ and zero suppressed with a threshold of \ODZeroSuppressionThreshold.  Veto data are recorded in a \ODAcquisitionWindow\ window beginning \SI{10.5}{\us} before the initiating TPC trigger~\cite{Agnes:2016cp}.

\subsection{Reconstruction}
\label{sec:Reconstruction}

Low-level reconstruction of \TPC\ events follows the steps described in Ref.~\cite{Agnes:2015gu}.  The digitized \PMT\ waveforms that make up the raw data are analyzed using \DSfDarkart, a code based on the Fermi National Accelerator Laboratory {\bf art} framework~\cite{Green:2012do}, which identifies pulses with area $\gtrsim$\DSfDdPulsefinderThr\ in the acquisition window.   Timing and integral information are calculated for each pulse.  While pulse-finding is done on the veto data, the WIMP search uses only integrals over pre-specified regions of interest (ROI), described in \ref{sec:NeutronCuts}. 

Of particular interest is the TPC \PSD\ parameter \FNine, defined as the fraction of \SOne\ light detected in the first \WindowFNine\ of a pulse.  This parameter allows very strong pulse shape discrimination between \NR\ and \ER~\cite{Boulay:2006hu}, as demonstrated in practice in our previous work~\cite{Agnes:2015gu}. 

In the current analysis, we perform radial fiducialization using transverse ($x$-$y$) reconstruction.  We did not do so in previous \DSf\ analyses, as reconstruction of the $x$-$y$ position of events in \DSf\ proved to be very difficult~\cite{Brodsky:2015uw,Watson:2017wl}.  This is believed to be due to the proximity of the top \PMTs\ to the \STwo-emission region, which limits charge sharing among the \DSfPMTSize\ \PMTs.  The ($x$-$y$) position reconstruction algorithm used here~\cite{Brodsky:2015uw} starts with maps constructed from Monte Carlo events of the simulated light response for each PMT vs.~the true position.  
Atmospheric argon data~\cite{Agnes:2015gu}, dominated by uniformly-distributed $^{39}$Ar decays, are used to iterate the maps to account for features in data not modeled in the Monte Carlo.  For a given event in the data, the algorithm compares the measured pattern of S2 light with the maps, finding the position that gives the best agreement.  
The position resolution is estimated to be about \SI{0.6}{\cm} from the observed spatial separation of events tagged as a delayed coincidence of \ce{^214Bi}-\ce{^214Po} decays.
These events were distributed across the full volume and had an \STwo\ signal size of about \SI{20E3}{\pe}.  
In the absence of any internal calibration sources with known location, we found no reliable way to calibrate the resolution vs.~the absolute position.  We discuss how we dealt with this situation to estimate the rejection and acceptance of the radial cut in Secs.~\ref{sec:ER} and \ref{sec:Acceptance}.

After reconstruction, data are stored in a ROOT format~\cite{Brun:1997ce}.  This is summarized in a secondary output called SLAD (for SLim Analysis Data), with event and pulse information for further study by analyzers.  Separate SLAD are made for the \TPC\ and veto data.  These are then matched event-by-event using the timestamps in each data stream.

\subsection{Calibration}
\label{sec:Calibration}

The single-photoelectron (\SPE) response of each \PMT\ in the \TPC\ and vetoes is determined by injecting low-light-level laser pulses into the detector volumes via optical fibers.  The \SPE\ means and widths are determined in the \TPC\ and vetoes as described in ~\cite{Saldanha:2017eq,Agnes:2015gu,Agnes:2016cp}.

The \SOne\ light yield is measured using \ce{^{83m}Kr} introduced into the recirculating argon~\cite{Kastens:2009hz}.  The \ce{^{83m}Kr}  decays to \ce{^{83}Kr} in two sequential transitions, where the second transition has a mean-life of \KrEightThreeMTwoMeanLife\ and thus is usually reconstructed as part of S1.  This provides a monoenergetic signal in the \TPC\ that is also used to calibrate the \SOne\ signal $z$-dependence and the \STwo\ signal radial dependence as described in~\cite{Agnes:2015gu, Agnes:2016fz}.  The zero-field \UAr\ photoelectron yield at the \TPC\ center, measured at the \KrEightThreeQValue\ \ce{^{83m}Kr} peak, is \DSfDdNullFieldLightYield.  \ce{^{83m}Kr}  campaigns taken at various times during the running period indicate that it remained stable within \DSfDdNullFieldLYStability.  

While we use \SOne\ as our primary energy variable in the WIMP search, the sharing of deposited energy between scintillation and ionization in a \TPC\ makes a combination of \SOne\ and \STwo\ a more linear and higher-resolution energy variable~\cite{Pagani:2017uh,Hackett:2017vg}.  We use such a variable for the determination of background-generating radioactivity in the detector -- see Sec.~\ref{sec:Activity}.  The combined \SOne-\STwo\ \ER\ energy scale is established by reconstructing \gr\ lines from trace radioactivity in detector components.  These lines at higher energies consist of multiple Compton scattering events, requiring special techniques to deal with events with multiple \STwo\ pulses~\cite{Pagani:2017uh}.  

We construct the nuclear recoil energy scale from the \SOne\ signal using the photoelectron yield of \NRs\ of known energy measured in the \SCENE\ experiment~\cite{Alexander:2013ke,Cao:2015ks}, via the procedure described in Ref.~\cite{Agnes:2015gu}.
Briefly, \SCENE\ measures the ratio of \NR\ yield at \DSfDriftField\ to that of \ce{^{83m}Kr}  at zero field.  Our zero-field photoelectron yield for \ce{^{83m}Kr}  then gives the \NR\ PE yield vs.\ \SOne\ in \DSf.  We assume constant \NR\ PE yield above the highest \SCENE-measured energy, \SCENERecoilsEnergyMax.

Initial operations of \DSf\ with atmospheric argon (AAr)~\cite{Agnes:2015gu} provided a large sample of \ce{^{39}Ar} $\beta$~decays.  This data set of uniformly-distributed, single-sited ER events is used as our primary calibration of \FNine\ (see Sec.~\ref{sec:ER}).

Coincident \grsnh\ from \ce{^60Co} decays in the cryostat steel are used to determine the \LSV\ light yield and to measure the time-offset between the \TPC\ and \LSV\ signals.  Cosmic-ray muons align the timing of the \WCV\ with the other detectors.  

\AmBe\ neutron calibrations are used to determine the \FNine\ distribution for \NR.  \AmBe\ NR candidates are selected by requiring a single-sited \TPC\ event in prompt coincidence with an \LSV\ signal consistent with a \AmBeGammaEnergy\ $\gamma$~ray from the source.  The \FNine\ distributions for each \SOne\ bin in this data sample are fitted with an analytic model based on a modified ratio-of-two-Gaussians treatment~\cite{Edkins:2017vf, Agnes:2016tx}.  This analytic model is used only for calculating the acceptance of the final \FNine ~vs.~S1 WIMP search box (see Sec.~\ref{sec:Acceptance}). 

\AmBe\ calibrations are used to establish the heavily-quenched visible energy of the neutron captures on \ce{^10B} that give the \LSV\ its high efficiency for captures~\cite{Agnes:2016fw}.  
Coincident \grsnh\ preclude the use of \AmBe\ for calibrating the prompt neutron thermalization signal in the \LSV.  For this we use an \AmC\ source~\cite{Liu:2015ig} with a thin degrader that reduces the $\alpha$ energy below that needed to reach the lowest excited state of \ce{^16O}.  With lead shielding to absorb the \ce{^241Am} x-rays, this results in a neutron source very low in coincident \grsnh, allowing study of isolated neutrons (see Sec.~\ref{sec:Radiogenic}).

\subsection{Data Set}
\label{sec:DataSet}

The data set reported in this paper consists of \DSfDdLTPostQualCut\ of \UAr\ data taken from \DSfDdStartDate, to \DSfDdEndDate.  It does not include the data reported in Ref.~\cite{Agnes:2016fz}.  Aside from \TPC\ laser calibration runs (typically taken several times per day and lasting \DSfDdLaserrunDuration) and occasional calibration campaigns (\ce{^{83m}Kr}, \AmBe, and \AmC, lasting a few days to a few weeks), data were taken continuously in \DM-search mode, and running conditions were very uniform throughout this period.  Data were usually divided into runs of 6-hour duration.

The trigger rate varied from \SIrange{1.3}{1.7}{\Hz} due to intermittent bipolar noise spikes generated by a high-voltage power supply.  These spikes were completely removed by our baseline-finding software, leaving a residual rate of \SI{1.2}{\Hz} mostly due to \grsnh\ from detector materials and \ce{^85Kr} and residual \ce{^39Ar} in the \UAr~\cite{Agnes:2016fz}.

Blinded data (see Sec.~\ref{sec:Blinding}) were checked run-by-run for hardware and software issues that warranted run removal.  The main causes were oscillations in veto-channel front end electronics (\DSfDdLSVOscLoss) and abnormal baseline noise in \TPC\ \PMT\ signals (\DSfDdTPCPedRMSLoss), with smaller losses from runs shorter than \DSfDdMinRunEvt\ (\DSfDdMinRunTime\ duration), individual \TPC\ \PMTs\ breaking down or emitting light, and other causes.  After eliminating these runs, the total livetime of the data set was \DSfDdLTPostRV.  This is reduced further by event quality cuts and the veto cut against cosmic ray activation (see Sec.~\ref{sec:Studies} and Table~\ref{tab:AccSummary}) to our final livetime of \DSfDdLTPostQualCut.  With our fiducial mass of \DSfDdFiducialMass\ (see Sec.~\ref{sec:Acceptance}), the exposure reported here is \DSfDdExposure.

\section{Background Sources and Mitigation}
\label{sec:Background}

Processes that provide backgrounds to the \DM\ search fall into two main categories. The first category consists of $\alpha$ decays and neutrons, which yield \NR\ or \NR-like signals strongly resembling \DM\ scatters.  The second category consists of ER-inducing processes, primarily $\beta$~decays and $\gamma$-ray interactions, that, although more copious, are suppressed by the powerful PSD in \LAr.  In this section we describe the major background categories and our mitigation strategies.  The background rejection levels achieved and the  levels of background expected in the final sample after all cuts are given in Sec.~\ref{sec:Studies}.

\subsection{$\alpha$ Decays}
\label{sec:Alphas}

For $\alpha$ decays in the active \LAr, or on or very near surfaces touching it,  both the $\alpha$\ itself and the recoiling daughter nuclide give \NR-like \FNine. 
Given the highly radiopure materials selected for construction of the \TPC, the $\alpha$~emitters of interest are primarily radon daughters either deposited on detector surfaces during fabrication and assembly or introduced into the circulating \LAr\ during the experiment.

\begin{figure}[t!]
\centering
\includegraphics[width=0.5\textwidth]{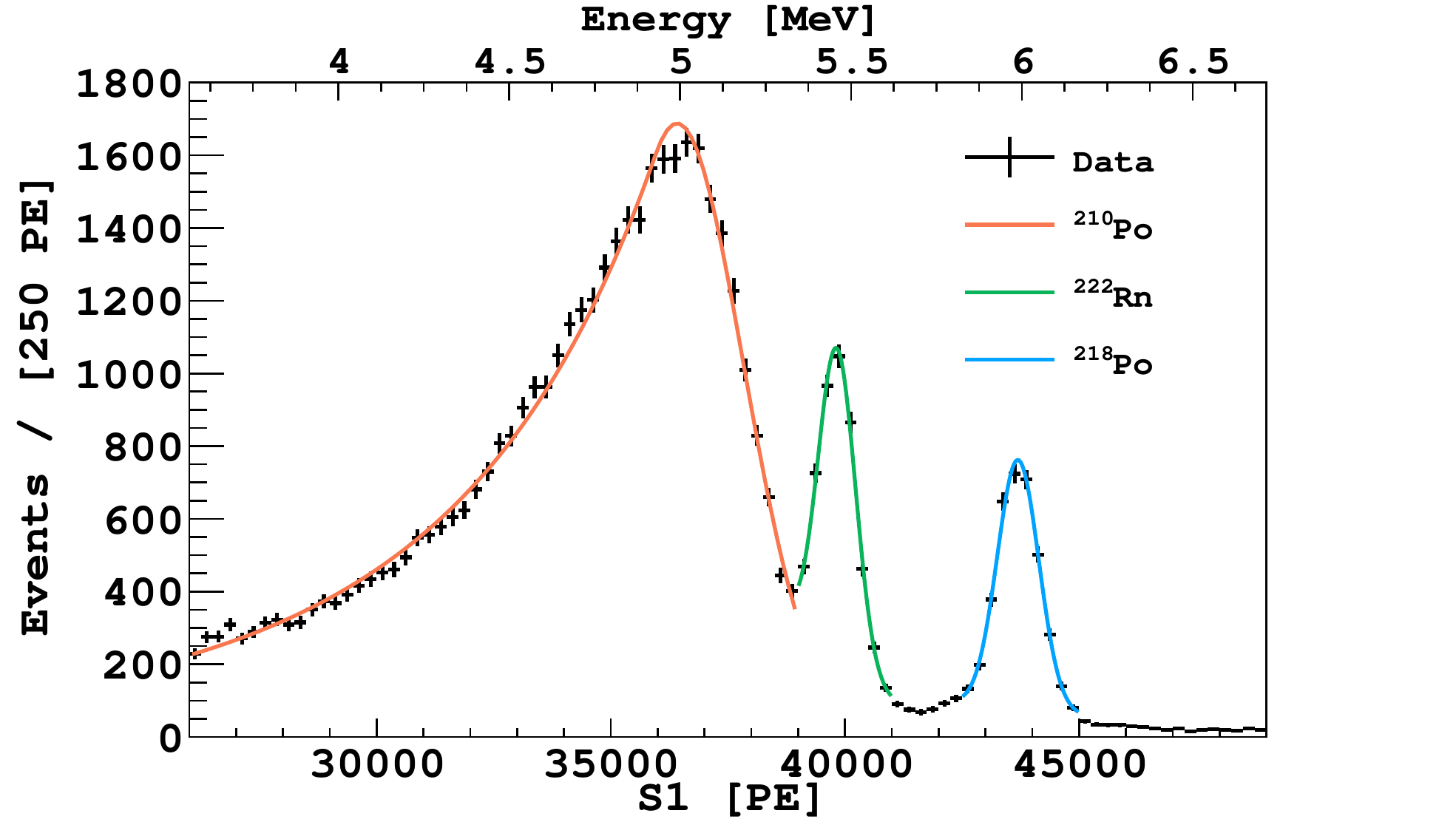}
\caption{$\alpha$ spectrum.  The events are selected by requiring the first pulse to be alpha-like ($\num{0.5} < \FNine < \num{0.9}$).  Two corrections are applied to \SOne.  The first corrects for \ADC\ saturation, and the second corrects for the $z$-dependence of the light yield~\cite{Agnes:2017cl}.  The \ce{^222Rn} and \ce{^218Po} peaks are fitted with Gaussian+exponential functions, and the energy scale at the top of the plot is set by the \ce{^218Po} peak~\cite{Stanford:2017wl}.  The \ce{^210Po} is fitted with a Crystal Ball function, the shape of which suggests that the \ce{^210Po} is on the surface beneath the \TPB.} 
\label{fig:AlphaSpectrum}
\end{figure}

We have seen and studied  both surface and bulk-\LAr\ $\alpha$ events in \DSf~\cite{Stanford:2017wl}, with an energy spectrum shown in Fig.~\ref{fig:AlphaSpectrum}.  The measured specific activities of \ce{^222Rn} and \ce{^218Po} in the LAr are ($2.12\pm 0.04$)~\si{\micro\becquerel\per\kilo\gram} and ($1.55\pm 0.03$)~\si{\micro\becquerel\per\kilo\gram}, respectively~\cite{Stanford:2017wl}.  $\alpha$~decays in the bulk \LAr\ give sharp peaks in \SOne\ which are far outside the \DM\ \NR\ energy range, leaving surface events to contend with as background.

The major source of surface background is \ce{^210Pb}-supported \ce{^210Po} decays.  With the full-energy \ce{^210Po} $\alpha$'s outside the \DM-search energy range, the potential background sources are either $\alpha$'s degraded in energy or events with the daughter \ce{^206Pb} atom recoiling directly into the \LAr.  The broad lineshape of the \ce{^210Po} signals identified in Ref.~\cite{Stanford:2017wl} and shown in Fig.~\ref{fig:AlphaSpectrum} gives clear evidence for degraded $\alpha$ events.  The recoiling atoms alone would not produce enough light in the \LAr\ to be a background, but simultaneous $\alpha$ scintillation in the \TPB\ can boost the event into the \DM\ search region~\cite{Xu:2017jy}.

Surface events on the cathode and grid are easily rejected by drift time cuts ($z$ fiducialization).  The observed rate of \ce{^210Po} $\alpha$'s on the $\sim$\SI{0.4}{\square\meter}~side reflector is \DSfDdPoTwoTenSideRate.  Section~\ref{sec:Surface} discusses several characteristics of surface events, beyond their radial location, that allow them to be rejected.

\subsection{Neutrons}
\label{sec:Neutrons} 

Individual elastic scatters of neutrons in the \LAr\ are indistinguishable from \DM-induced scatters, making these a critical background.  Considerable efforts in \DSf\ were devoted to reducing and suppressing neutron background, most notably stringent materials selection and the development of the veto system.  

Neutrons are produced by cosmic-ray muons interacting in the rock and other materials surrounding the experiment (cosmogenic) and by trace radioactivity of detector materials (radiogenic).  Many neutron-induced events can be rejected because, unlike \DM\ particles, the neutrons are very likely to interact multiple times in our detectors.  Multiple interactions in a single \TPC\ event are detected by resolving multiple \STwo\ pulses.  Both cosmogenic and radiogenic neutrons leaving WIMP-like signatures in the \TPC\ also leave signals in the \LSV\ with high probability, allowing them to be rejected with high efficiency.  (See Sec.~\ref{sec:Radiogenic} for details.)  Additional rejection in the \TPC\ comes from fiducialization (again due to the relatively short neutron interaction length in \LAr), and from requiring \SOne\ to lie in the WIMP search range.  The \WCV\ gives additional rejection of cosmogenic events.

Radiogenic neutrons come from spontaneous fission of \ce{^238U} and from \alphan\ interactions, where the $\alpha$'s come from uranium and thorium chain activity.  In \DSf, the spontaneous fission events are  easily rejected due to the high \LSV\ efficiency for neutrons and moderate efficiency for \grsnh, combined with the average neutron multiplicity for \ce{^238U} spontaneous fission of \UTwoThreeEightSFAvgNeutronMult\ and the high \gr\ multiplicity.  This leaves \alphan\ as the main source of potential radiogenic neutron background.

Our \alphan\ calculations~\cite{Westerdale:2017gx}, normalized to the assayed construction materials activities described in Sec.~\ref{sec:Activity}, indicate that the dominant sources of neutron production in the \TPC\ and cryostat are the \PMTs\ and a viton o-ring in the outer cryostat flange.  For neutrons that reach the \TPC\ and give single-scatter \NR-like events in the fiducial volume, \Geant-based Monte Carlo simulations (\GFDS)~\cite{Agnes:2017cz} indicate that the o-ring contribution is negligible, and the \PMTs, specifically the borosilicate-glass ``stem" and the ceramic plates that hold the dynodes, are the source of \SI{>90}{\percent} of the radiogenic neutron background in the \TPC.

\subsection{$\beta$ Decays and $\gamma$ Rays}
\label{sec:BetaGamma}

The \WCV\ and \LSV\ provide efficient passive shielding against $\beta$'s and \grsnh\ originating outside the \TPC\ cryostat,  leaving the cryostat and \TPC\ components (including the \LAr) as the only important sources of \bg-induced background.  Argon derived from the atmosphere (\AAr) contains $\sim$\AArArThreeNineActivity\ of cosmic-ray produced \ce{^39Ar} activity~\cite{Loosli:1983bu,Benetti:2007fg}.  \ce{^39Ar} is a $\beta$ emitter and dominated the trigger rate and background in \DSf\ when it was filled with \AAr~\cite{Agnes:2015gu}.  The DarkSide collaboration has identified, extracted, and purified argon from underground sources (\UAr)~\cite{AcostaKane:2008im, Back:2012vo, Back:2012um} that has only \DSfUArArThreeNineActivity\ of \ce{^39Ar} activity~\cite{Agnes:2016fz}.  The use of \UAr\  drastically reduces the ER background in \DSf.  Even including the \DSfDdKrEightFiveActivity\ of \ce{^85Kr} found in the current \DSf\ \UAr\ fill~\cite[corrected for the 15.5-y mean life]{Agnes:2016fz}, the dominant source of ER background is Compton scatters of \grsnh\ from the \TPC\ and cryostat.  

\PSD\ via \FNine\ is the major rejector of $\gamma$-induced \ER.  In Ref.~\cite{Agnes:2015gu} we showed that PSD with \FNine\ rejected the single-sited \ER\ events from \ce{^39Ar} decay to a level of one in \DSfAArROIEventsNumber.  Unlike the ER events from \ce{^39Ar}, $\gamma$-induced events are often multi-sited and are not uniformly distributed, so requiring single-scatter events and fiducializing give additional suppression.  Many $\gamma$-induced events in the \TPC\ are in prompt coincidence with additional interactions in the LSV, giving further rejection. 

The fundamental limitation on \PSD\ removing single-sited \ER\ scintillation events is at low energies, where photoelectron statistics limit rejection.  
However, among \gr-induced events, there are some in which a $\gamma$~ray multiple-Compton scatters, scattering once in the active LAr and also in a nearby Cherenkov radiator such as the Teflon reflector or the fused silica windows of the TPC or PMTs.  The all-prompt Cherenkov light adds to the prompt component of the normal \ER-like \SOne\ and can give a \NR-like \FNine.  As discussed in Sec.~\ref{sec:ER}, these mixed scintillation+Cherenkov events, hints of which had already appeared in Ref.~\cite{Agnes:2016fz}, prove to be the dominant background in the experiment.

\subsection{Determination of Activities in Detector Materials}
\label{sec:Activity}

The \gr- and neutron-induced backgrounds originate  primarily in the trace radioactivity of detector components.  The \DS\ collaboration carried out an extensive program of assays to select radiopure materials and to understand their residual activities.  Our background estimates are based on a radioactivity model that starts with the results of the assays.  However, due to a late-developing need to use R11065 PMTs instead of the planned lower-activity R11065-20s, we do not have assays of the PMTs installed in \DSf, but rather only a single measurement of three R11065s from early production batches.

For this reason, activities in the PMTs are estimated by fitting spectra generated by Monte Carlo from activities in various detector locations to a reconstructed TPC energy spectrum~\cite{Koh:2018un,Pagani:2017uh}. Since the actual construction materials used for the cryostat components (stainless steel body, flanges, nuts, bolts, pipes/feedthroughs, Viton o-ring, multi-layer insulation) were assayed, their respective activities in the fitting process are fixed to the assayed values. The $^{39}$Ar and $^{85}$Kr in the LAr are fixed to their values as reported in \cite{Agnes:2016fz}, with the $^{85}$Kr corrected for its decay since that measurement.

We consider the activities of these isotopes in the PMTs: $^{60}$Co, $^{40}$K, $^{232}$Th, $^{235}$U, and $^{238}$U 
(allowing secular equilibrium to be broken, with $^{226}$Ra as the top of the lower chain).
The main hosts of radioactivity in the PMTs are the borosilicate glass stem at the back of the PMT, the ceramic insulators supporting the dynodes, and the Kovar casing.  Comparing the results of assays of the ceramic insulators, a Kovar casing, and various versions of whole R11065 PMTs, the fraction of each activity in each PMT component was inferred, and we fit the summed PMT activities keeping these fractions fixed.

\begin{figure}[t!]
\centering
\includegraphics[width=0.5\textwidth]{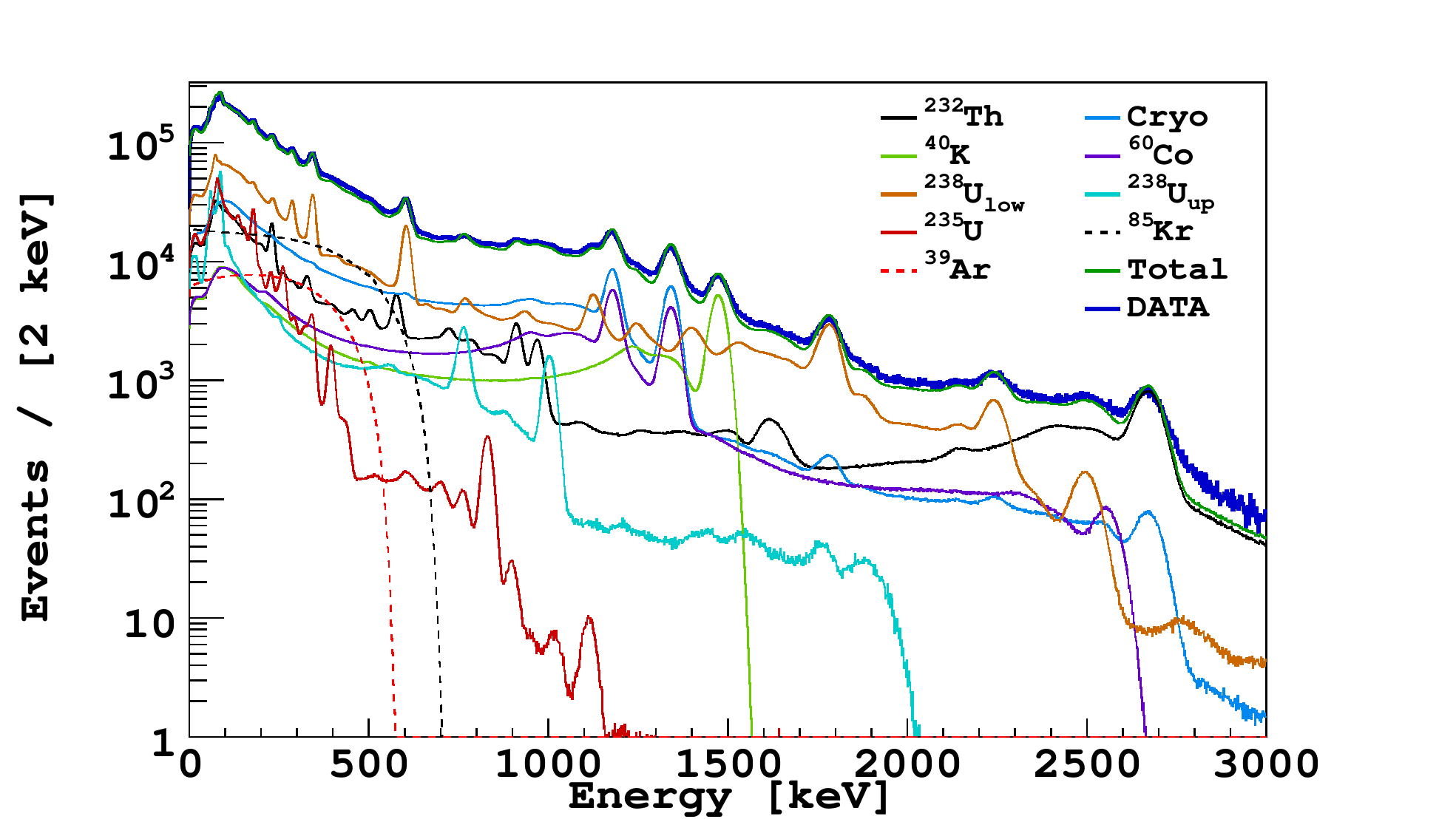}
\caption{Measured \gr\ spectrum in the \TPC\ (dark green) with the total from the fit (dark blue) including cryostat activity (light blue) fixed to assayed values and fitted \PMT\ activities (see legend). The energy scale is the combined S1-S2 ER energy scale (see Sec.~\ref{sec:Calibration}).} 
\label{fig:BackgroundSpectra}
\end{figure}

The fit is done iteratively, estimating the PMT activities by taking advantage of certain high-energy \grsnh\ unique to individual decay chains.  $^{232}$Th activity in the PMTs is estimated first by fitting the 2.6~MeV $^{208}$Tl peak, where the contribution from the other decay chains is low. $^{232}$Th activity is then fixed at the fitted best value, and the $^{238}$U lower chain ($^{238}$U$^{\text{low}}$) activity is estimated by fitting the 1.76~MeV $^{214}$Bi peak, and so on. The $^{235}$U and the $^{238}$U upper chain ($^{238}$U$^{\text{up}}$) activities are fitted with one free parameter to preserve their natural abundance ratio. The activity estimates from this procedure are presented in Table~\ref{tab:Activities} and the resulting energy spectrum is shown in Fig.~\ref{fig:BackgroundSpectra}. We note that leaving $^{85}$Kr and $^{39}$Ar free in the fit along with $^{235}$U and $^{238}$U$^{\text{up}}$ returns significantly different rate estimates for these four decay chains; however, switching between the rates so-obtained and those presented in Table~\ref{tab:Activities} has no impact on the predicted background in the WIMP search region.  Note as well that, while the WIMP-search region is far to the left in Fig.~\ref{fig:BackgroundSpectra}, the thorium and lower uranium chains, fitted to the right side of the plot, are the main contributors to Cherenkov radiation, from electrons scattered by the high energy \grsnh, and neutrons, produced by high energy $\alpha$'s.

The uncertainty on the PMT background activity from a given chain is estimated by propagating the uncertainty on the measured cryostat activity in that chain. (The uncertainties from the fit are negligible.)  In particular, the uncertainties on $^{60}$Co, $^{40}$K, $^{232}$Th, and $^{238}$U$^{\text{low}}$, the main contributors to Cherenkov background due to their high energy \grsnh, are estimated to be $<$13\%.

\begin{table}
\centering
\caption{TPC component activities, estimated by fitting $^{232}$Th$_{\text{PMT}}$, $^{238}$U$^{\text{low}}_{\text{PMT}}$, $^{40}$K$_{\text{PMT}}$, and $^{60}$Co$_{\text{PMT}}$ in sequence, followed by $^{235}$U$_{\text{PMT}}$ and $^{238}$U$^{\text{up}}_{\text{PMT}}$ while $^{85}$Kr and $^{39}$Ar are fixed at their measured rates as reported in \cite{Agnes:2016fz}. Cryostat activities  are fixed at their measured rates from assays and summed across all cryostat locations. PMT activities are summed across all locations within the PMTs and across all 38 PMTs.  For comparison, we show the assayed activities for 3 R11065 PMTs (scaled to 38 PMTs), which have an estimated additional systematic uncertainty of about 25\%.  
}
\label{tab:Activities}
\begin{tabular*}{0.9\columnwidth}{c|c|c||c}
\hline \hline
Source								&\multicolumn{2}{c||}{PMTs [\si{\becquerel}]}	&Cryostat [\si{\becquerel}]\\
									&fitted			&assayed				&assayed\\
\hline 
$^{232}$Th				&0.277$\pm$0.005		&0.23$\pm$0.04			&0.19$\pm$0.04 \\ 
$^{40}$K				&2.74$\pm$0.06			&3.0$\pm$0.4			&0.16$^{+0.02}_{-0.05}$ \\ 
$^{60}$Co				&0.15$\pm$0.02			&0.17$\pm$0.02			&1.4$\pm$0.1 \\ 
$^{238}$U$^{\text{low}}$	&0.84$\pm$0.03			&0.69$\pm$0.05			&0.378$^{+0.04}_{-0.1}$ \\ 
$^{238}$U$^{\text{up}}$	&4.2$\pm$0.6				&5.3$\pm$1.1			&1.3$^{+0.2}_{-0.6}$ \\ 
$^{235}$U				&0.19$\pm$0.02			&0.27$\pm$0.4		&0.045$^{+0.007}_{-0.02}$ \\ 
\hline
								\multicolumn{4}{c}{Liquid Argon Activity [mBq/kg] }   \\
\hline
$^{85}$Kr							&\DSfDdKrEightFiveActivityNum &$^{39}$Ar							&0.7$\pm$0.1 \\ 
\hline \hline
\end{tabular*}
\end{table}

\section{Blinding Scheme}
\label{sec:Blinding}

We performed a blind analysis on the \DSfDdLTPostQualCutnum-live-day data set.  This means that candidate selection/background rejection was designed, and the background surviving cuts was estimated, without knowledge of the number or properties of events in the final search region.  

\begin{figure}[t!]
\centering
\includegraphics[width=0.5\textwidth]{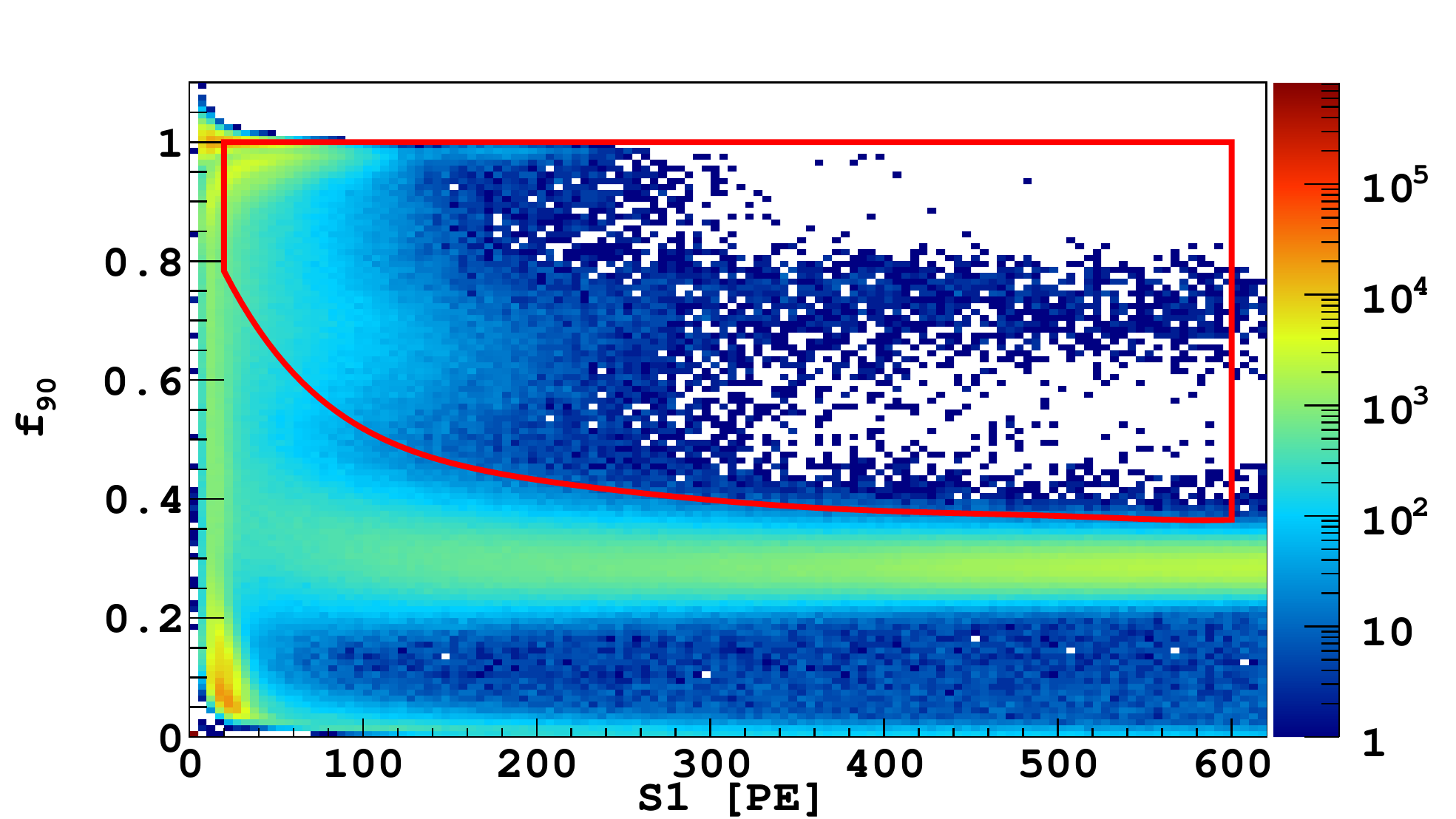}
\caption{\FNine ~vs.~\SOne\ showing the blinding box (red) applied to the Ref.~\cite{Agnes:2016fz} data set.} 
\label{fig:DSfDdBlinding}
\end{figure}

Blindness was imposed by a ``Blinding Module'' in SLAD.  
An unblinded SLAD was produced first and kept in a protected directory.  Then the SLAD program operated on it with the Blinding Module to produce the blinded, analyzer's version.
Blinded events appear in the output files, but with all \TPC\ data except the event ID, timestamps, and the livetime associated with the event set to \num{-1}.  
In the initial blinding, used through most of the analysis, details of two categories of events were hidden from users. The first category consisted of events with \SOne\ and \FNine\ falling within the ``blinding box", shown in Fig.~\ref{fig:DSfDdBlinding} superimposed on the published data set from Ref.~\cite{Agnes:2016fz} before any analysis cuts.  The blinding box was designed to be larger than any expected final WIMP-search box and to be just above the main ER band.  It was applied to all events, even those that failed major analysis cuts (e.g., single-pulse events, events with multiple \STwo's, etc.).  
The second category consisted of events randomly chosen with a probability of \DSfDdRanBlindProb.
The random fraction was chosen to have enough fluctuations to obscure the counting of possible candidate events in the final analysis stages, where it was anticipated that the number of candidates would be small or zero when final cuts were applied.

Besides the events outside the blinding box, open data available to analyzers included the large \AAr\ data set~\cite{Agnes:2015gu}, the initial \DSfUArAfterQualityCutsLiveTimeAdj\ \UAr\ data set~\cite{Agnes:2016fz},  laser calibration data, and all data from campaigns with calibration sources present.  
During the analysis, we opened sections of the blinded data outside of the WIMP search region to provide samples enriched in particular backgrounds for study, and later, when the background predictions were mature, to test the predictions.
Several such test regions, described below, were studied before the final box opening. 

\section{Background Estimation and Rejection}
\label{sec:Studies}

The goal of the blind analysis is to design a set of criteria that rejects background to a pre-determined level without prior inspection of events in the final search region (the ``box"), which itself must be designed as part of the analysis procedure.  We choose 0.1 event of expected background as an acceptable level, giving a $<$10\% Poisson probability of seeing one or more background events in the search box.

\subsection{Event selection}
\label{sec:Cuts}

As in earlier \DSf\ analyses, the initially-dominant ER background and the power of LAr PSD suggest an analysis structured around the \FNine ~vs.~\SOne\ distribution.  We thus choose the design of the \FNine ~vs.~\SOne\ box as the final analysis step, after all other cuts are defined.

We began with the set of analysis cuts developed for earlier analyses~\cite{Agnes:2015gu,Agnes:2016fz}.   Some of these cuts were modified for this analysis, and some new ones were developed -- the new or modified cuts are indicated with asterisks.  We introduce all the cuts here with brief descriptions; the full set is listed in the acceptance table, Table~\ref{tab:AccSummary}.  The motivations for some of the cuts will be elaborated on in the sections describing the relevant backgrounds.

\subsubsection{Event quality cuts}
\label{sec:qualcuts}

\noindent{\sf AllChan}: data are present for all TPC channels in the event.

\noindent{\sf Baseline}: baselines for the digitized waveforms are successfully found in all TPC channels.

\noindent{\sf VetoPresent}: the event has GPS-timestamp-matched veto data.

\noindent{\sf TimePrev*}: the event occurs at least \DSfDdMinLivetimeCut\ 
after the end of the inhibit window of the previous trigger (that is, at least \SI{1.21}{\milli\second}  
after the previous trigger).  This removes events that triggered on an \STwo\ whose \SOne\ occurred during the inhibit window.

\subsubsection{Basic TPC event cuts}
\label{sec:basiccuts}

These cuts are designed to ensure that passing events are single-scatter events that triggered on \SOne\ and have a single valid \STwo.

\noindent{\sf S1start}:  the first pulse occurs at trigger time. 

\noindent{\sf Npulse}:  there is a second pulse, presumed to be \STwo.  A third pulse is allowed only if its timing is consistent with the small tertiary pulses produced when \STwo\ light photoionizes the TPC cathode. 

\noindent{\sf S1sat}:  the first pulse does not saturate the digitizers.

\noindent{\sf MinS2uncorr*}: the second pulse is required to be $\ge$\DSfDdSTwoUncorrMin\ before position-based corrections, the approximate threshold for successful reconstruction of the event's radial position.  For reference, the uncorrected S2's of interest in this analysis are $>$\DSfDdMinSTwouncorrActual.

\noindent{\sf S2f90}: the second pulse has \FNine$<$\DSfDdSTwoFnMin, consistent with the slow rise-time of \STwo\ pulses.

\noindent{\sf xyRecon}: the $x$-$y$ reconstruction algorithm successfully derives transverse coordinates of the event from \STwo.

\noindent{\sf MinS2/S1}:  a more refined S2 cut that removes events with unphysically small \STwo/\SOne.  The cut is set to remove events in the lowest 1\%\ of the S2/S1 distribution of \AmBe\ NRs.

\subsubsection{Surface background cuts}

These cuts were all developed for the current analysis~\cite{Stanford:2017wl}.  They are described in Sec.~\ref{sec:Surface}.

\noindent{\sf LongS1tail*}:  removes events with \SOne\ with a long tail, consistent with laboratory measurements of $\alpha$-induced scintillation in TPB wavelength shifter.  

\noindent{\sf MaxS2/S1*}: removes events in the highest 1\%\ of the S2/S1 distribution of \AmBe\ NRs.  This cut targets the ``Type 2" surface background with uncorrelated \SOne\ and \STwo\ described in Sec.~\ref{sec:Surface}.  This can also be a powerful discriminant between NR and ER and is the basis of WIMP discrimination in LXe TPCs.  In \LArTPC's it is effective against high-energy ERs, but it is not effective at low S1, where further rejection is most needed.  

\noindent{\sf S2LEshape*}: removes events in which the shape of the leading edge of the second pulse is not consistent with the shape of a true \STwo\ pulse~\cite{Agnes:2018dt}.

\noindent{\sf S1TBA*}: removes events with a $z$ location determined from the \SOne\ top-bottom asymmetry that is not consistent with the $z$ location determined from \STwo\ via $t_{\rm drift}$.

\subsubsection{Neutron background cuts}
\label{sec:NeutronCuts}

The neutron veto cuts are essentially unchanged from the first \UAr\ analysis~\cite{Agnes:2016fz,Agnes:2016fw}.

\noindent{\sf LSVprompt}: rejects events with $>$\DSfDdLSVpromptThreshold\ in the interval \DSfDdLSVpromptInterval\ relative to the TPC trigger time.  This targets the thermalization signal from neutrons giving NR in the TPC.

\noindent{\sf LSVdelayed}: rejects events with  $>$\CutHVnewdellongpemax\ in a \CutHVnewsliderwidth\ sliding window covering \DSfDdLSVdelayedInterval\ after a TPC trigger.  This interval can be compared to the capture lifetime of \LSVNewNeutronCaptureMeanLife\ in the boron-loaded liquid scintillator.  The long acquisition window and search interval allow us to veto efficiently via the emitted \grsnh\ even when the neutron captures in TPC materials with long capture lifetimes.

\noindent{\sf LSVpre}: rejects events with $>$\pvcutpemax\ in a \CutHVnewsliderwidth\ sliding window covering \DSfDdLSVprepromptInterval\ before a TPC interaction. 

\noindent{\sf CosmicMu}: rejects events with a \WCV\ signal \mvcutwcvpemax\ or an LSV signal \mvcutlsvpemax, integrated over the full \ODAcquisitionWindow\ acquisition window. This vetos cosmic-ray muons or their showers and thus cosmogenic neutrons. 

\noindent{\sf CosmoActiv*}: a``cosmic ray activation veto" is applied if a TPC event occurs within \DSfDdWCVmuonActInterval\ (shorter than in previous analyses) following a triggered event failing the {\sf CosmicMu} cut.  This removes some delayed neutrons produced by cosmic-ray-activated isotopes in the detectors. 

\subsubsection{ER background cuts}

PSD via \FNine\ is the primary discriminant against ER backgrounds and is used to define the final WIMP search box via the procedure discussed in Sec.~\ref{sec:ER}.  We found in this analysis that scintillation+Cherenkov events dominated the tail of the \FNine\ distribution near the WIMP search region.
They thus determined the search box needed to reduce the total background to $<$0.1 event in the full exposure.  The other cuts aimed at scintillation+Cherenkov and other ER events are discussed here.

\noindent{\sf tdrift}: vertical fiducialization via the time between \SOne\ and \STwo\ ($t_{\rm drift}$) is effective against \grsnh\ from the PMTs, their primary source.  We use the same vertical fiducialization as in the previous analyses, removing \DSfDdtdriftFid\ of drift time ($\sim$\DSfDdtdriftFidcm) from the top and bottom of the active volume.  Though the ER background determined the location of the cut, it is also clearly important for surface background, notably serving to eliminate $\alpha$ decays occurring on the TPC cathode and grid.

\noindent{\sf S1pMaxFrac*}:  for ``S1 prompt maximum fraction", removes events with \SOne\ too concentrated in any one PMT.  These events are likely to have interactions giving Cherenkov light in the fused silica PMT and TPC windows.  A variant of this cut was used in past analyses, but it was modified for the current analysis to use only prompt light, boosting its effectiveness as a Cherenkov discriminant.  This cut is extremely effective against fused silica Cherenkov, leaving scintillation+Cherenkov in the Teflon reflector as the main surviving ER background.

\noindent{\sf S1NLL*}:  squeezes further rejection from the \SOne\ PMT pattern, targeting the multi-sited nature of scintillation+Teflon Cherenkov events.  The pattern of \SOne\ light on the PMT arrays is required to be consistent with the reconstructed $x$-$y$ position via a negative-log-likelihood comparison to templates derived from AAr data (dominated by single-sited $^{39}$Ar $\beta$ decays).

\noindent{\sf RadialFid*}: a radial fiducial cut.
The radial cut is a drift-time-dependent radial contour chosen to reject a fixed fraction of \GFDS-simulated scintillation+Teflon Cherenkov events (see Sec.~\ref{sec:ER})  in each drift time bin.  The final cut varies from $\sim$\DSfDdRadCutTopBot\ from the wall at the top and bottom of the TPC to $\sim$\DSfDdRadCutCenter\ from the wall at the center. 
 
\begin{figure}[t!]
\centering
\includegraphics[width=0.5\textwidth]{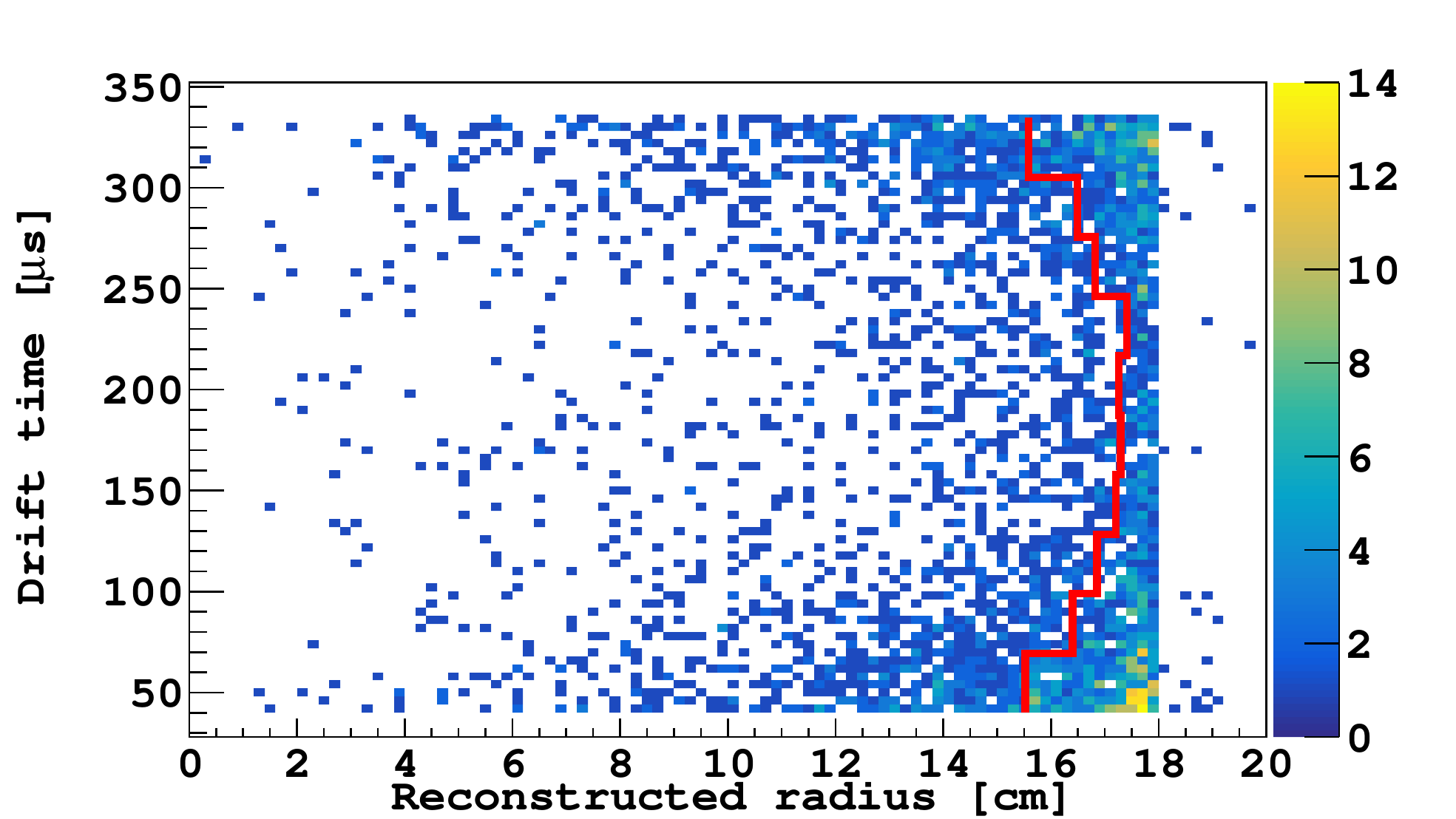}
\caption{Radial cut (red) shown on events in the initial blinding box shown in Fig.~\ref{fig:DSfDdBlinding}.  
All event quality cuts (Sec.~\ref{sec:qualcuts}), all basic cuts (Sec.~\ref{sec:basiccuts}) through {\sf MinS2uncorr}, and vertical fiducialization via $t_{\rm drift}$ have been applied.} 
\label{fig:DSfDdRadialCut}
\end{figure}

The effect of the radial cut is shown in Fig.~\ref{fig:DSfDdRadialCut}, made after unblinding.  The events (primarily ER background from PMT and cryostat \grsnh, including mixed scintillation+Cherenkov events) are seen to be concentrated near the top and sides of the detector as expected. Despite the limitations of the  reconstruction algorithm, the concentration of events and the impact of the cut are clear.

\subsection{Surface events}
\label{sec:Surface}

Alphas coming from isotopes embedded in detector surfaces exhibit a degraded energy spectrum and can fall within the energy and $f_{90}$ regions of interest, as can the recoiling nucleus in an $\alpha$ decay~\cite{Xu:2017jy}.
We find that the S2 signal for surface events in \DSf\ is heavily suppressed, possibly 
due to loss of drifting electrons very close to the side reflector of the TPC. Few surface events have an S2 that is large enough to pass analysis cuts, with the majority having no discernible S2 pulse. We call these ``S1-only" events.

We therefore consider two cases for a surface decay to become a background event. Type 1: the rare case of a surface event with a true S2 that passes analysis cuts. Type 2: an S1-only event that happens to occur before an uncorrelated ``S2-only" event such that the combination appears to be a regular event with one S1 and one S2. 
We estimate the background rates of these two cases separately.

Type 1: In the open data with S1$>$600 PE, surface events only pass S2 analysis cuts at energies far above the region of interest (S1$>$20\,000 PE) due to the low electron collection efficiency along the side wall, with an acceptance that declines with decreasing S1. Extrapolating this effect into the WIMP search region and applying it to the observed rate of S1-only events, we estimate that \DSfDdSurfBkgndPassSTwo\ such events could pass the S2 analysis cuts.

Further reduction of Type 1 surface background is achieved by using the layer of TPB deposited on detector surfaces as a veto. It was discovered that alphas passing through TPB induce millisecond-long scintillation in the wavelength shifter~\cite{Stanford:2018un}. The presence of this slow component following an S1 pulse tags the event as originating from a TPB-coated surface. 
We count the individual photoelectrons in the region between S1 and S2 and define a cut based on this count, {\sf LongS1tail}, that accepts 99\% of \AmBe\ NR events. Applying the cut to a sample of surface decays obtained in Ref.~\cite{Stanford:2018un} results in a rejection factor of more than 100, giving an expectation of \DSfDdSurvBkgndSurfTypeOneMod\ Type 1 surface background events in the current data set. Additional rejection is expected from the {\sf RadialFid} and {\sf MinS2uncorr} cuts, which is difficult to estimate and not included in the background estimate. 

Type 2:  True S2-only events are rare, but apparent S2-only events are present in the form of ordinary events near the top of the detector. In these events, S1 and S2 can be so close in time ($t_{\rm drift}\lesssim$\DSfDdMinResSOneTwoTdrift) that they are not resolved by our reconstruction. 
The real or apparent S2-only events and S1-only surface events are uncorrelated and of constant rate, allowing the use of Poisson statistics to predict the expected number of S1+S2 pileup background events.

We mitigate Type 2 background by imposing three additional requirements on the apparent S2 signal. 
The first is the maximum S2/S1 cut, {\sf MaxS2/S1}, which removes events with S2/S1 larger than 99\%  of \AmBe\ NR events of the same S1.  This cut targets S1-only events with an accidental S2 either augmented by an unresolved S1 or simply uncorrelated with S1. 
The second, {\sf S2LEshape}, removes unresolved S1 and S2 by requiring that the apparent S2 pulse have the $\sim$\SI{2}{\micro\second} risetime of a true S2 pulse~\cite{Agnes:2018dt} rather than the few-ns risetime of S1.
This S2 shape cut is applied via the ratio of the integrals of the first 90~ns and first 1$\mu$s of the S2 pulse. 
The third, {\sf S1TBA}, removes events with S1 and S2 pulses that originate from different positions.  
We require that the $z$ positions inferred from the top-bottom asymmetry in the detected S1 light and from $t_{\rm drift}$ differ by no more than 3$\sigma$, as determined from uniform $^{39}$Ar events from AAr. These last two cuts are each designed to have $>$99\% acceptance for nuclear recoils. After application of these additional cuts, we expect \DSfDdSurvBkgndSurfTypeTwo\ 
surviving Type 2 surface background events in the current data set.

\subsection{Radiogenic neutrons}
\label{sec:Radiogenic}

The estimate of radiogenic neutron background starts with a direct measurement of the LSV efficiency for detecting neutrons that leave WIMP-like signatures in the TPC.  We do this with the \AmC\ source (see Sec.~\ref{sec:Calibration}) deployed just outside the TPC cryostat.  
\AmC\ calibration data are taken in the same trigger configuration as normal WIMP-search data, with the TPC triggering both vetoes.
The standard WIMP analysis is run to find NR candidates in a preliminary version of the \FNine~vs.~\SOne\ WIMP-search box.  The neutron veto efficiency is then calculated as the fraction of TPC NR candidates that fail the standard WIMP-search LSV cuts described in Sec.~\ref{sec:NeutronCuts}.  From a sample of about 25,000 events that pass TPC NR cuts, we find the veto efficiencies shown in Table~\ref{tab:AmCeff}. 
Radiogenic-neutron background events differ from \AmC -neutron events in their origin point and energy spectrum, but  Monte-Carlo simulations indicate a {\it higher\/} veto efficiency for radiogenic events;  we do not apply that correction here.

\begin{table}
\centering
\caption{Neutron veto efficiencies for \AmC\ source data.  Errors are statistical.  The prompt cut targets neutron thermalization; the delayed cut neutron capture.
}
\label{tab:AmCeff}
\begin{tabular}{c | c | c}
\hline \hline
Prompt cut only			& Delayed cut only		& 	Combined	\\ \hline 
\DSfDdLSVEffPrompt	& \DSfDdLSVEffDelayed	&  \DSfDdLSVEffCombined		\\ 
\hline \hline
\end{tabular}
\end{table}

One of the test regions opened prior to the final unblinding was the ``Veto Prompt Tag" (VPT) sample, which unblinded any event that failed the {\sf LSVprompt} cut.  The high neutron efficiency of the prompt cut allows radiogenic neutron events to be counted directly in the VPT sample. 
The narrow integration window of the LSV-prompt cut means that, even with its \DSfDdLSVpromptThreshold\ threshold, the accidental tagging rate is $<$1\% (see Table~\ref{tab:AccSummary}).  Thus the VPT tag accidentally accepts practically no real WIMP events, and $\gamma$-induced events are the only background to a neutron count using the VPT sample. 

To get a sample of confirmed neutron events from the VPT sample, we use a modified version of the LSV-delayed cut.  The modification is needed for two reasons: the sliding window used for the LSV-delayed cut overlaps the LSV-prompt window (albeit with a higher threshold), and the LSV has a high rate of PMT afterpulses, so the delayed region is heavily populated by afterpulses from the prompt signal.
The modified LSV-delayed cut uses LSV cluster-finding~\cite{Agnes:2016fw} to identify veto hits.  To count as a likely neutron capture signal, the cluster is required to be $>$\DSfDdVPTdelayedStart\ after the veto prompt time, to have the number of PMTs contributing to the cluster greater than that expected for afterpulses, and to have an integral $>$\DSfDdVPTCaptThr, which includes the $\alpha+\gamma$ capture peak for \ce{^{10}B}~\cite{Agnes:2016fw} and captures on \ce{^{1}H} and \ce{^{12}C}.  
The neutron efficiency for this restricted capture signal is calculated from \AmC\ data to be
$\sim$\DSfDdVPTneutronEff, with most of the inefficiency coming from exclusion of the \ce{^{10}B} $\alpha$-only capture peak and 
events from the $\alpha+\gamma$ capture peak in which the $\gamma$~ray escapes into the cryostat.
(None of these complications or efficiency losses apply to the actual neutron vetoing in the WIMP search, which is done with simple integrals over regions-of-interest -- see Sec.~\ref{sec:NeutronCuts}.) 

\begin{figure}[t!]
\centering
\includegraphics[width=0.5\textwidth]{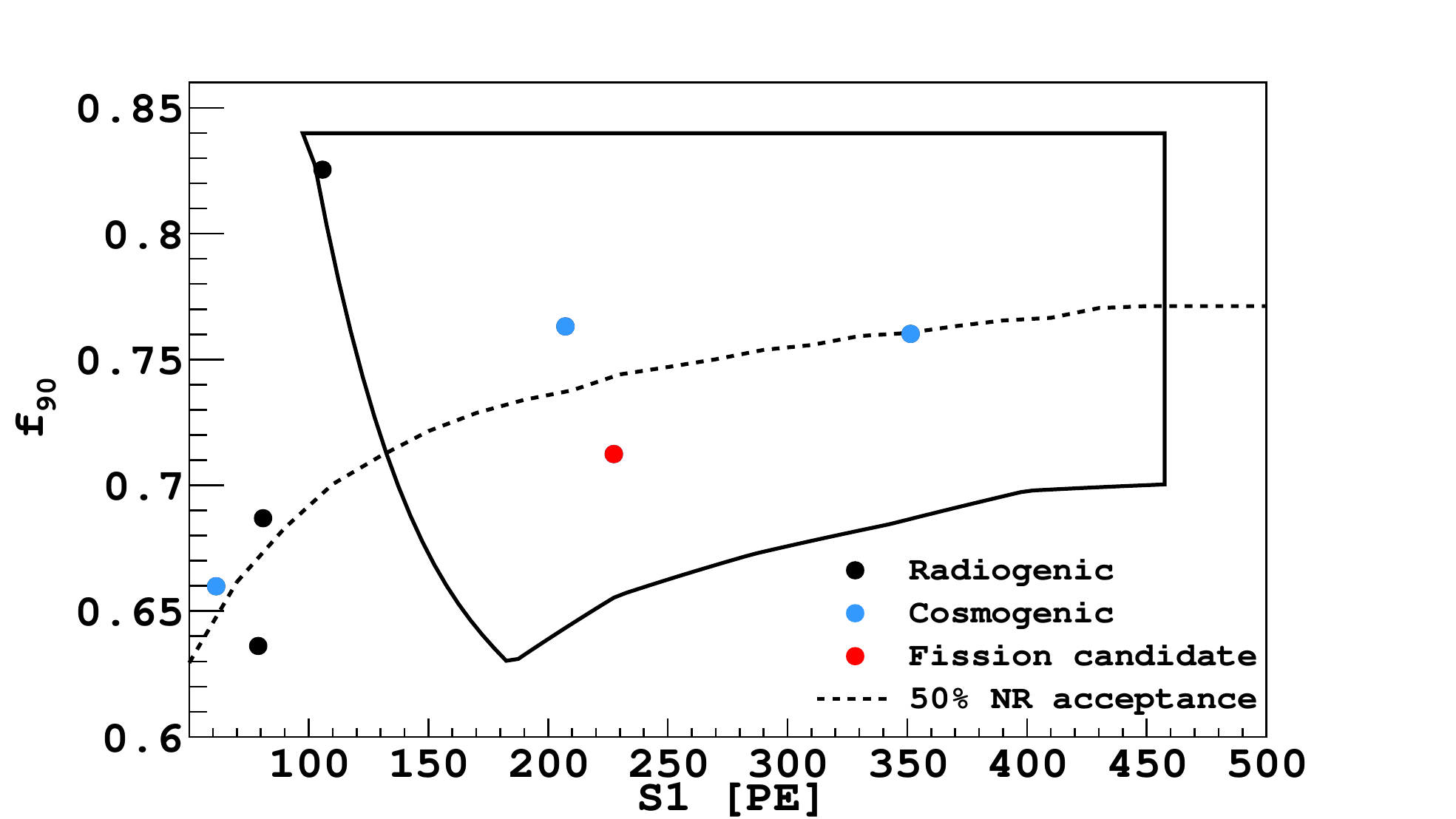}
\caption{Neutron candidates in the Veto Prompt Tag sample.  The closed curve is the final WIMP-search box, and the dashed curve is the 50\% NR contour, around which neutron-induced events should be distributed.} 
\label{fig:DSfDdNeutrons}
\end{figure}

The selected neutron candidate events are shown in Fig.~\ref{fig:DSfDdNeutrons}, where we label neutron candidates that fail the {\sf CosmicMu} cut as ``Cosmogenic", and a spectacular event with three neutron capture signals as ``Fission".  
There is one observed radiogenic \alphan\ neutron in the WIMP-search region in the VPT sample.  With an acceptance of \DSfDdVPTneutronEff\ for the neutron counting and a veto efficiency greater than 
$\epsilon_{AmC}^{data} = \DSfDdLSVEffCombined$, we predict a radiogenic neutron background \DSfDdSurvBkgndRadioNeutron\ events, with \SI{100}{\percent} statistical error.

\subsection{Cosmogenic Neutrons}
\label{sec:Cosmogenic}

The rate of cosmogenic neutron background is estimated via simulation using
FLUKA (version 2011.2c)~\cite{Bohlen:2014dv,Ferrari:2005fa}. The simulation is carried out in multiple steps. 
In the first and most time-consuming step, cosmic-ray muons are started 7~m above the ceiling of LNGS Hall C and propagated through the 7 m of rock.  The muon and any produced secondaries are stopped and stored when they reach the ceiling of Hall C \cite{Empl:2014ih}. 
The stored events are restarted and propagated onto the \WCV\ and are only processed further if there are no muons entering the water tank with energy $>$\DSfDdCosmoMuMaxE\ and projected path length in the water $>$\DSfDdCosmoMuMaxPath, since these would be rejected by the \WCV. We find that for a generated livetime of \DSfDdCosmoMCGen, the FLUKA simulation predicts \DSfDdCosmoMCreachTPC\ events in which any particle reaches the TPC.  None of the \DSfDdCosmoMCreachTPC\ events passes the simulated veto cuts.  
Only one event is a single neutron in the TPC with no other accompanying particles.
In 6 more events, a neutron is accompanied by one other particle that is not an easily rejected muon, typically a \gr\ or another neutron.  None of these 7 events have TPC energy deposits in our WIMP-search region.  

If we take a 90\% CL upper limit of 2.3 of \DSfDdCosmoMCreachTPC\ events reaching the TPC passing the veto cuts and take the 7 (neutron+$\leq$1 particle) events as a conservative upper limit on the number of neutron events passing TPC cuts in \DSfDdCosmoMCGen, we predict \DSfDdSurvBkgndCosmoNeutron\ cosmogenic neutron events passing all cuts in the present WIMP search.

When we include the muons with long path lengths and high energies in the \WCV, the rate of simulated single-scatter neutron events in the TPC  depositing energy in the WIMP search region rises to $\sim$2 per year, in agreement with our count of 3 cosmogenic neutrons, shown in Fig.~\ref{fig:DSfDdNeutrons}.  

\subsection{Electron Recoil Backgrounds}
\label{sec:ER}

With the PSD performance demonstrated in the AAr run of \DSf~\cite{Agnes:2015gu} and the reduced rate from the use of UAr~\cite{Agnes:2016fz}, the most tenacious ER background is mixed scintillation+Cherenkov events.
To estimate ER background surviving cuts, a data/MC hybrid model was developed, which incorporates our GEANT4 simulation to model the $\gamma$-ray kinematics and Cherenkov radiation while drawing \FNine\ from the AAr data. 

A very large sample of Monte Carlo simulated events, equivalent to about 90 live-years of data, was generated. Statistics this large were needed to ensure that 0.05~events of ER background in our exposure would be represented by at least 3 Monte Carlo events. This was chosen so that, based on the 68\% C.L. interval constructed in~\cite{Feldman:1998bu}, the statistical uncertainty on the background prediction would be no more than a factor of two. Events were generated representing the decay chains and TPC components listed in Table~\ref{tab:Activities}.  These were later normalized to the activities in that Table and the accumulated livetime of the WIMP-search data.  
To save on computation time, S1 photons for individual LAr scatters are typically generated but not tracked. However, for events with Cherenkov radiation, all photons -- including those from LAr scintillation, if there is an accompanying scatter in the LAr -- are generated and tracked using optical parameters tuned on data~\cite{Agnes:2017cz}.

Cherenkov light can be generated in the fused silica PMT windows, the fused silica TPC windows, and the Teflon reflectors surrounding the active LAr volume.  The optical parameters affecting the Cherenkov radiation and collection are adjusted to match the observed ``pure Cherenkov" events in data, which are easily identified as single-pulse events with $\FNine\!\approx\! 1.0$ (all prompt light). 
A high-statistics sample of pure Cherenkov events, enriched in events with Cherenkov light generated in the Teflon side reflector, was obtained using a 11.2~kBq $^{22}$Na source deployed next to the TPC cryostat.   The modeling of the generation of Cherenkov photons and their collection by the PMTs was subsequently validated against pure Cherenkov events from the open UAr data set and from the $^{241}$Am-Be calibration data, and the scintillation+Cherenkov background model was frozen.

\begin{figure}
\centering
\includegraphics[width=0.5\textwidth]{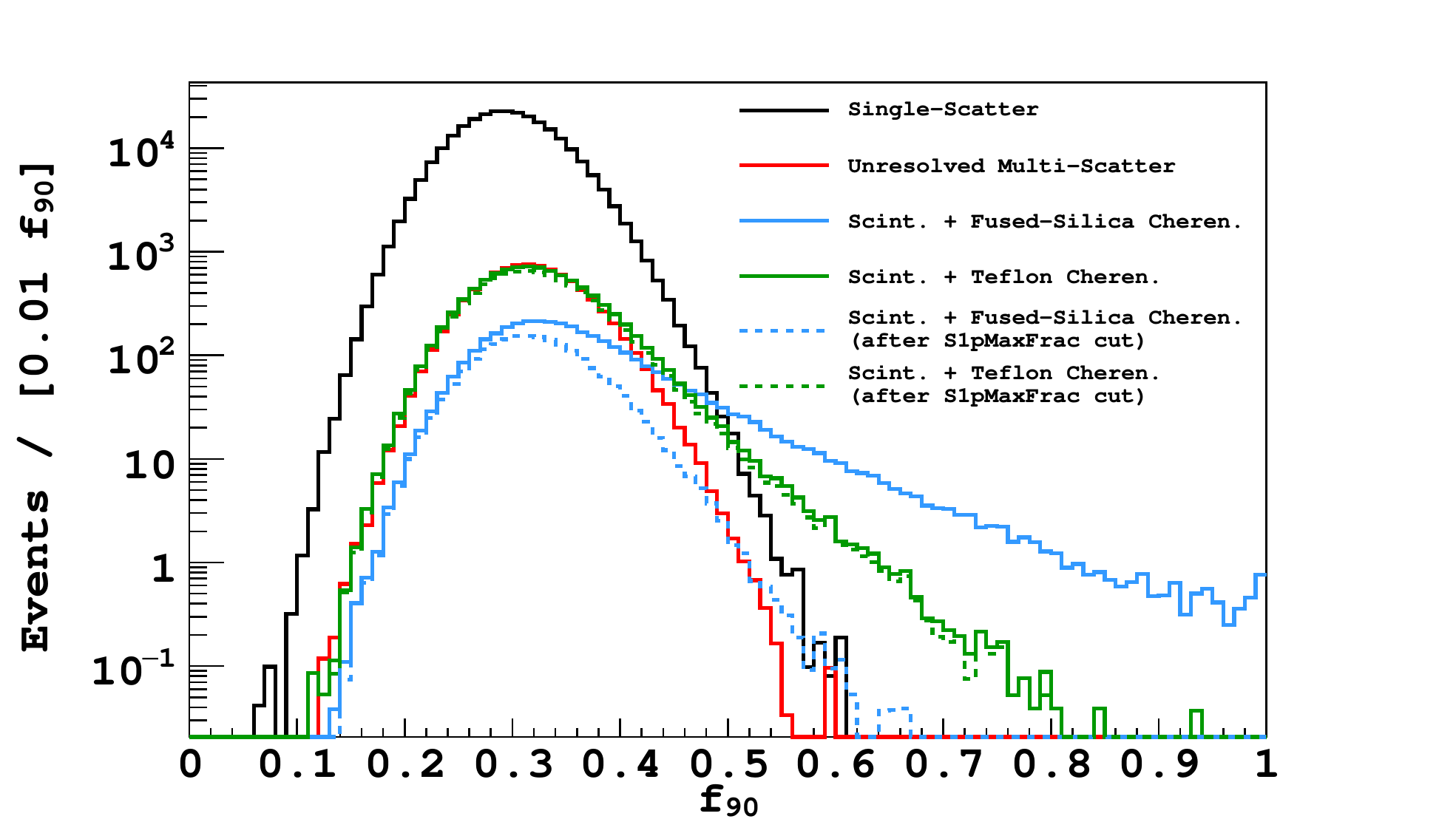}
\caption{Modeled \FNine\ profiles of single-scatter, unresolved multiple-scatter, scintillation+fused silica Cherenkov, and scintillation+Teflon Cherenkov 2-pulse events with 100$<$S1$<$180~PE. Decay chains and activities in the various detector locations in Table~\ref{tab:Activities} are used. It is clear that the {\sf S1pMaxFrac} cut is very effective on high \FNine\ events with a FS Cherenkov component (blue), hence the most problematic background comes from Teflon Cherenkov.}
\label{fig:PSDModel}
\end{figure}

The model constructs the \FNine\ of a simulated multiple-scatter event from the \FNine's of its component scatters.  Energy depositions in the LAr with a vertical separation $<$\DSfDdSTwoMinSep\ (motivated by studies using our electronics simulation) are merged to model our S2 two-pulse resolution. Figure~\ref{fig:PSDModel} shows that \FNine\ for unresolved multiple-scatter events is higher than that of single-scatters with the same S1, since mean ER \FNine\ increases with decreasing S1. We estimate that unresolved multiple-scatters are 3\% of ER events with $100<\text{S1}<180$ PE  (the region where this is estimated to have the most impact).

With a targeted background level of $<$0.1~event in \DSfDdLTPostQualCut\ of data, we require reliable predictions far out in the tail of the ER \FNine\ distribution. For this analysis, we do not extrapolate using an analytic model fit to data. 
Instead, we use our high-statistics AAr data set~\cite{Agnes:2015gu}, which is dominated by uniformly distributed, single-sited ERs from \ce{^{39}Ar} $\beta$-decays, as the \FNine\ probability distribution function. 
In particular, modeled single-scatter events in the LAr (\ce{^{39}Ar} and \ce{^{85}Kr} $\beta$-decays and single Compton scatters of \grsnh) draw directly from the AAr \FNine~vs.~\SOne\ distribution, unresolved multiple Compton scatter events draw multiple times, and scintillation+Cherenkov events have their scintillation \SOne\ and \FNine\ augmented with the Cherenkov light predicted by the \GFDS\ model, treating the Cherenkov radiation as entirely prompt. The available AAr statistics, which represent about 15~years-worth of single-scatter events in UAr running, are sufficient, given the randomization that occurs when the scintillation \FNine's are combined with Cherenkov light.
 
ER background with Cherenkov light radiated in the fused silica PMT and TPC windows results in abnormally large amounts of light concentrated in individual PMTs. As shown in Fig.~\ref{fig:PSDModel}, the S1 prompt maximum fraction cut, {\sf S1pMaxFrac}, is very effective against fused-silica Cherenkov, leaving Cherenkov in the Teflon, primarily the cylindrical side wall of the TPC, as the dominant ER background.  

Attempts to find cuts effective against scintillation+Teflon Cherenkov events were only modestly successful.  
A major motivation for introducing a radial fiducial cut was its observed impact on high-\FNine\ events in the open data, as discussed below. 

Some cuts are difficult or impossible to apply to modeled events, so their impact in the search region is hard to estimate.  These include cuts based on S2, which was too costly in computation time to fully simulate in the large Monte Carlo sample, and cuts based on detector foibles like the surface background cuts discussed in Sec.~\ref{sec:Surface}.   Although they are applied to the data, we do not include potential rejection from the \STwo/\SOne\ cut, the NLL cut, and the surface background cuts in our ER background estimate.

Final testing of the model was carried out by unblinding various test samples.  These samples are of two types.  The first type consists of samples created by inverting an established analysis cut, giving events already tagged as background.  The second type consists of regions with small WIMP acceptance, outside any plausible final \FNine ~vs.~S1 box, but inside our (initially generous) blinding box.

\begin{figure}
\centering
\includegraphics[width=0.5\textwidth]{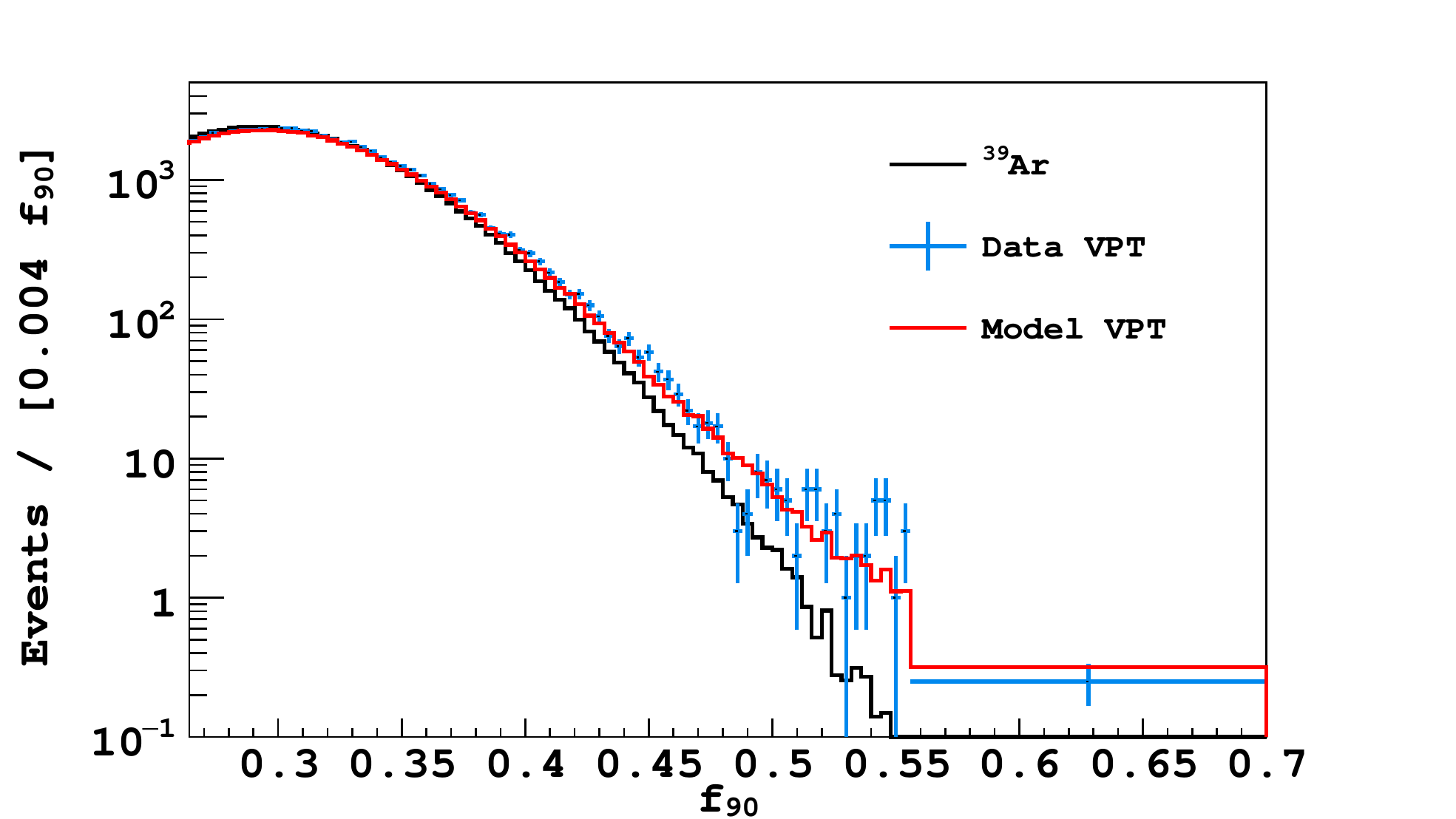}
\caption{\FNine\  for events with 100$<$S1$<$180~PE comparing single-scatter ER data (\ce{^{39}Ar}), VPT data passing all major cuts, notably {\sf S1pMaxFrac}, and simulated VPT events using the \FNine\ model including Cherenkov light.  The normalization of the simulation is absolute, using the activities in Table~\ref{tab:Activities} and the \DSfDdLTPostQualCut\ exposure.
}
\label{fig:DataModelVPT}
\end{figure}

The first of these tests uses the Veto Prompt Tag sample described in Sec.~\ref{sec:Radiogenic}.  
The number of neutrons in this sample was found to be small, and they are identified and removed.
While the VPT allows us to look in the WIMP search region without compromising blindness, the statistics are low.  We instead test the model in pre-defined regions near the search region with higher statistics, still dominated by scintillation+Teflon Cherenkov events.  Agreement in number of events between the data and model in these regions is within two statistical standard deviations.  Figure~\ref{fig:DataModelVPT} shows the data-model agreement in the \FNine\ spectrum.

\begin{figure}
\centering
\includegraphics[width=0.5\textwidth]{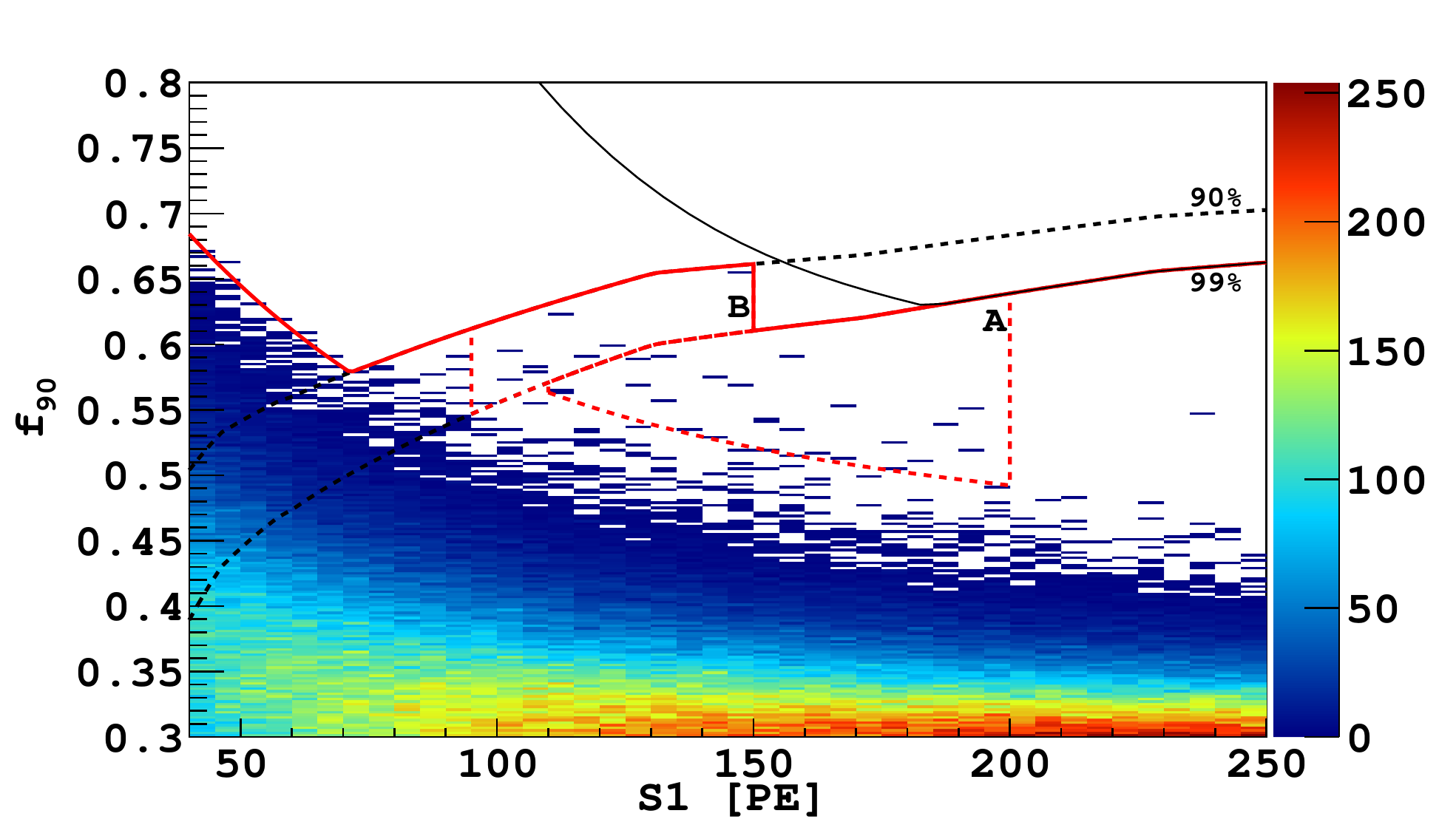}
\includegraphics[width=0.5\textwidth]{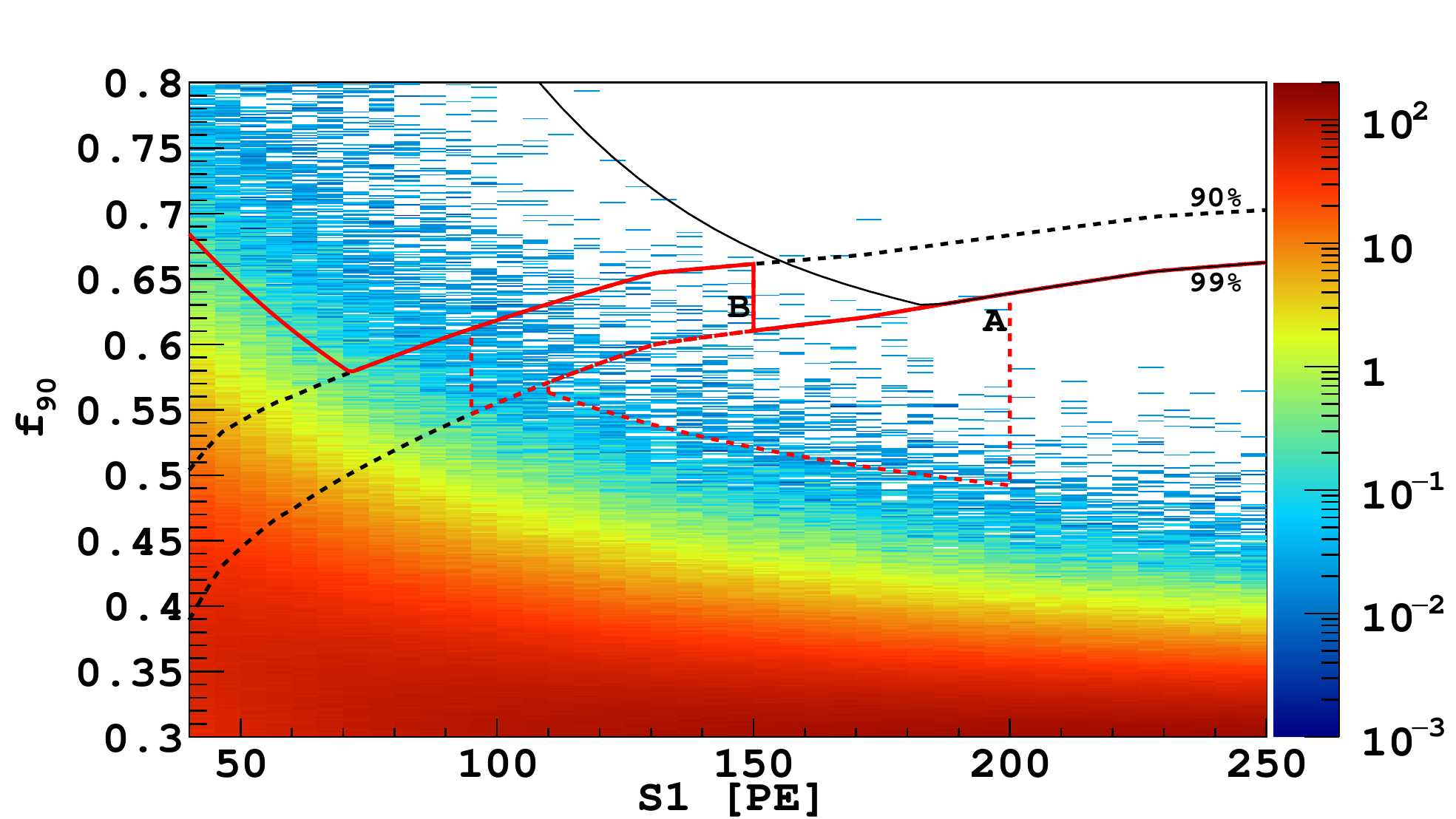}
\caption{\FNine ~vs.~S1 for events (top: data, bottom: model) passing all modeled cuts (see text). Overlaid on the plot are the two final test regions (dashed red), the remaining blinded region (solid red, and of course hidden in the data plot), and, for reference, NR 90\% and 99\% acceptance contours (dashed black) and the final WIMP search box (solid black).   Note: in the bottom plot, the high Monte Carlo statistics means that individual events have very small weights after normalization to data livetime, as reflected in the color axis of the plot. 
}
\label{fig:Triangle}
\end{figure}

In the final tests, we open two regions in the \FNine ~vs.~S1 plot, regions A and B in Fig.~\ref{fig:Triangle}, after applying the {\sf LSVprompt} cut and
all the TPC cuts other than the radial, NLL, S2/S1, and surface background cuts.
From the model, we expect these regions to be dominated by scintillation+Teflon Cherenkov events.
Table~\ref{tab:Tests} shows the data-model comparisons for the two test regions, with the model normalized to the data livetime. 
The first to be opened was region A. As can be seen in Table~\ref{tab:Tests}, the data and model in region A disagree at about the three-standard-deviation level statistically.  

Region B was designed and opened shortly after observing the data/model discrepancy in region A to supplement the available data statistics.  As can be seen in Table~\ref{tab:Tests}, the observed and predicted region B event counts are in agreement, albeit with poor statistics.

We choose to combine the statistics in regions A and B, and interpret the observed discrepancy between the data and model---a factor of 1.5---as a measure of the model's systematic error. Accordingly, we scale the model's output up by the same factor when making our ER background prediction.

The observed data events in regions A and B are also used to estimate the rejection of our radial fiducialization. After the remaining cuts not used in Table~\ref{tab:Tests} are applied, there are \DSfDdRegionABBeforeRadCut\ events, of which \DSfDdRegionABAfterRadCut\ survive the radial cut.  From this a rejection factor of (2.3$\pm$0.5) is inferred and applied to the model's prediction.  Hence, to design the final WIMP search box, we multiply the model background by $1.5/2.3$.

\begin{table}
\centering
\caption{Observed and predicted event counts in test regions A and B, shown in Fig.~\ref{fig:Triangle}. Note that while the model is normalized to the same livetime as the data, the model has vastly higher statistics, and thus negligible statistical errors.
}
\label{tab:Tests}
\begin{tabular}{l | c c}
\hline \hline
		& Region A 		& Region B  \\
		& events			& events \\
\hline
Data		& 24		& 9 \\
Model	& 13.3	& 8.7 \\
\hline \hline
\end{tabular}
\end{table}

Consistent with previous analyses, we fix the right edge of 
the WIMP box at S1=460~PE.
The presence of pure Cherenkov events in the data suggests that having a search box extending all the way to $\FNine=1$, as in past analyses, is unnecessarily risky, so we choose to put the upper edge of the box at $\FNine=\DSfDdBoxTop$, which is approximately the contour that excludes 1\% of NR.  At high S1, the background studies support a lower \FNine\ boundary than used in previous analyses -- we fix it along a curve that our latest calculations show to be approximately the 99\% NR acceptance contour.  
 At low S1, the box's lower boundary is determined by the desired total predicted background in the box (0.1~events in our case), and in particular, the ER backgrounds. With near-final estimates of the other backgrounds in hand, we allocated 0.08~background events to the ER backgrounds; the corresponding lower box boundary is drawn according to this requirement. 

In previous DarkSide analyses~\cite{Agnes:2015gu,Agnes:2016fz}, analytical models of \FNine\ fluctuations were fit to data in bins of S1, and the resulting functions were used to set a boundary that admitted equal background in each bin.  Adding Cherenkov light to the mix invalidates that procedure.  We use the ER background model described above for this purpose, but we do not have adequate Monte Carlo statistics for bin-by-bin assessment.  Instead, the determination of the boundary is done in two steps. 1) The rough shape of the boundary is determined where Monte Carlo statistics are available, by finding the \FNine\ that gives 0.07 leakage events in each 5~PE bin, about 14 times the final target background. A polynomial is fit to these points. 2) The fitted curve is translated upward in \FNine\ until the box defined by its intersection with the other bounds contains $\leq$0.08 events of ER background. In practice this was driven by 7 Monte Carlo events, to which we attached an uncertainty of $\pm$50\% by the construction in \cite{Feldman:1998bu}. This is the dominant uncertainty on the predicted ER and total background estimates.

\subsection{Background Summary and Cut Acceptance}
\label{sec:Acceptance}

A summary of the predicted backgrounds surviving all cuts in the full exposure is given in Table~\ref{tab:BkgndFinal}.  

\begin{table}
\centering
\caption{Predicted backgrounds surviving all cuts.  The ER background includes the scintillation+Cherenkov background.  The \FNine ~vs.~\SOne\ search box is defined to give \DSfDdSurvBkgndER\ surviving ER background events. 
}
\label{tab:BkgndFinal}
\begin{tabular}{l|c}
\hline \hline
Background			&Events surviving all cuts\\ \hline 
Surface Type 1		&\DSfDdSurvBkgndSurfTypeOneMod\\
Surface Type 2		&\DSfDdSurvBkgndSurfTypeTwo\\
Radiogenic neutrons	&\DSfDdSurvBkgndRadioNeutron\\
Cosmogenic neutrons	&\DSfDdSurvBkgndCosmoNeutron\\ 
Electron recoil		&\DSfDdSurvBkgndER\\ \hline
Total				&\DSfDdSurvBkgndTot\\ 
\hline \hline
\end{tabular}
\end{table}

\begin{figure}[t!]
\centering
\includegraphics[width=0.5\textwidth]{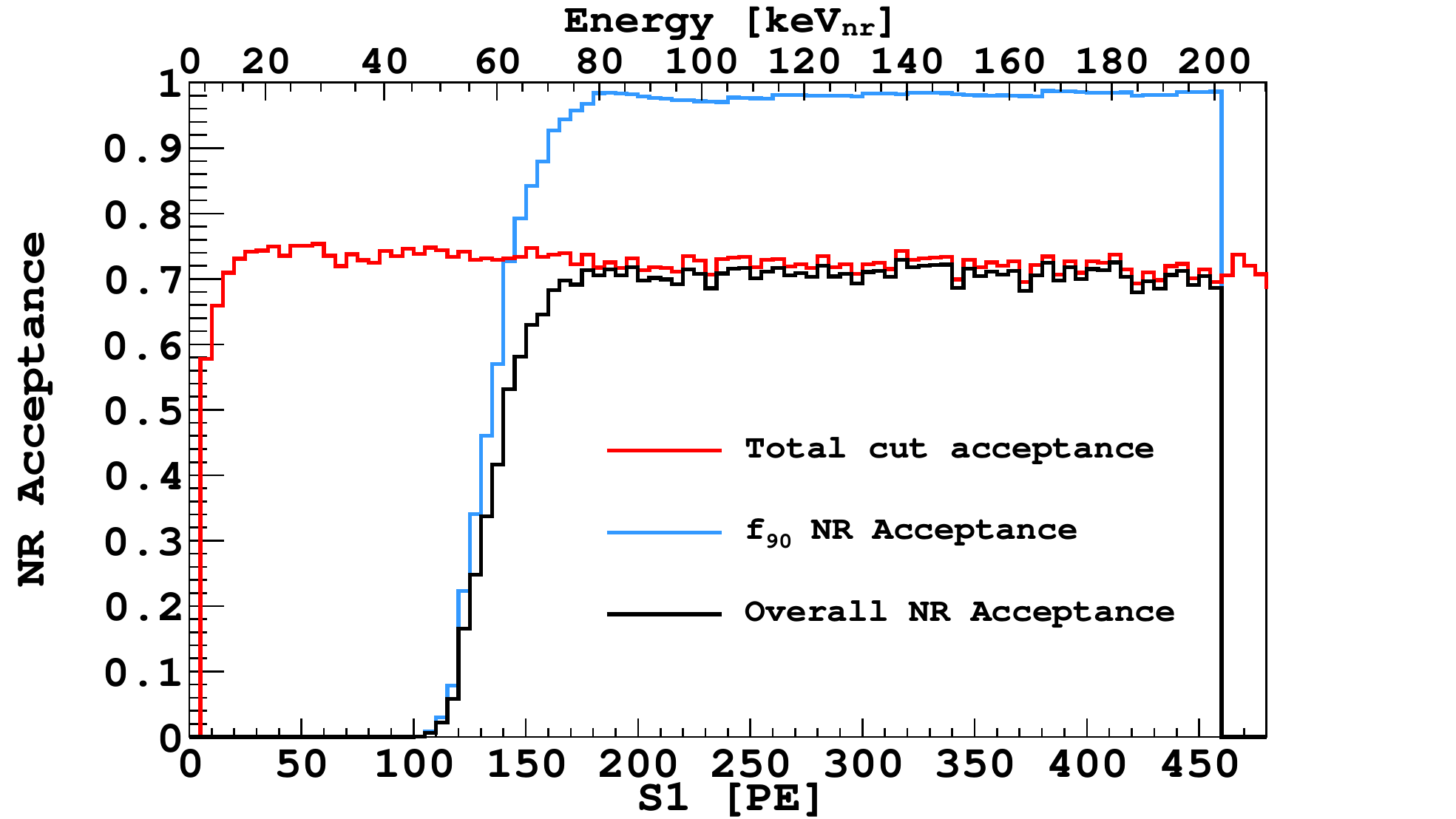}
\caption{Acceptance vs.~\SOne.  The \NR\ Energy scale at the top comes from the cross-calibration with \SCENE\ described in Sec.~\ref{sec:Calibration}.} 
\label{fig:Acceptance}
\end{figure}

The acceptance for each cut in the analysis except the fiducial cuts and the final \FNine~vs.~S1 WIMP search box is given in Table~\ref{tab:AccSummary}~\cite{Koh:2018un}.
With the exception of purely accidental losses such as those from the veto cuts, acceptances are measured with NR events from the \AmBe\ calibration data, corrected for spatial non-uniformity when necessary.  Several of the cuts have non-negligible S1 dependence.  In these cases, the full S1-dependent acceptance (see Fig.~\ref{fig:Acceptance}) is used to calculate the sensitivity of the analysis, and the Table-\ref{tab:AccSummary} entry is an average value.

\begin{table}
\small
\centering
\caption{Summary of cuts and their respective impact on livetime and WIMP acceptance.  The average acceptance of S1-dependent cuts are presented; acceptances $>$0.999 are shown as 1.  The cumulative acceptance is for all cuts except the fiducial and \FNine\ cuts.  Events surviving after each cut were tabulated after unblinding.
}
\label{tab:AccSummary}
\begin{tabular}{l | c | c}
\hline \hline
Cut 				& Livetime 		& \\
 				& (cumulative) 		& \\
\hline
{\sf AllChan}		& 545.6~d 		& \\
{\sf Baseline}		& 545.6~d 		& \\
{\sf TimePrev}		& 545.3~d 		& \\
{\sf VetoPresent} 	& 536.6~d 		& \\
{\sf CosmoActiv}	& 532.4~d 		& \\ 
\hline\hline
Cut				& Acceptance 		& Surviving events\\ 
				& (individual)		& in WIMP box\\ 
\hline
{\sf S1start}		& 1	 			& \\
{\sf S1sat} 		& 1	 			&  41884 \\
{\sf Npulse}		& 0.978 	 		& 726 \\
{\sf tdrift}			& (fiducial mass) 	& 191 \\
{\sf S1pMaxFrac} 	& 0.948 	 		& 21 \\
{\sf MinS2uncorr} 	& 0.996	 		& 4 \\
{\sf xyRecon} 		& 0.997	 		& 4 \\
{\sf S2f90}			& 1 	 			& 4 \\
{\sf MinS2/S1}	& 0.995	 			& 4 \\
{\sf MaxS2/S1}	& 0.991	 			& 4 \\
{\sf S2LEshape}	& 1	 			& 4 \\
{\sf S1TBA}		& 0.998 	 		& 4 \\
{\sf LongS1tail}		& 0.987 	 		& 3 \\
{\sf S1NLL}		& 0.99 	 		& 3 \\
{\sf RadialFid}		& (fiducial mass)	& 2 \\
{\sf CosmicMu}		& 0.990 	 		& 2 \\
{\sf LSVprompt} 	& 0.995			& 0 \\
{\sf LSVdelayed} 	& 0.835	 		& 0 \\
{\sf LSVpre}		& 0.992 	 		& 0 \\
\hline
Cumulative\rule[-5pt]{0cm}{15pt}		& $0.725\pm 0.001\;{\rm (stat)}$ \\
								& $^{+0.005}_{-0.004}\;{\rm (syst)}$ \\
\hline \hline
\end{tabular}
\end{table}

The impact of the fiducial cuts on sensitivity are counted in the fiducial mass.
The effect of the {\sf tdrift} cut, unchanged from previous analyses, is calculated from the geometry and drift velocity.  
The acceptance of the {\sf RadialFid} cut (see Fig.~\ref{fig:DSfDdRadialCut}) requires special treatment because of our lack of an absolute calibration for the $x$-$y$ reconstruction and because it is in principle S1 dependent via the S2-dependent $x$-$y$ resolution. We use the fact that $^{39}$Ar events are uniformly distributed like WIMP scatters and \AmBe\ events have \NR\ S2/S1 like WIMP scatters to determine the acceptance in two steps. 1) The cut's acceptance vs.~S2 is estimated using $^{39}$Ar events in our AAr data, which are uniformly distributed. 2) Acceptance vs.~NR S1 is then estimated by using S2/S1 as measured in our $^{241}$AmBe data to look up acceptance in the corresponding AAr S2 bin. Averaged over S1 in the WIMP selection region, the acceptance of this cut (after the drift time fiducialization) is \DSfDdRcutAcceptance, varying by less than 0.5\% with S1.  (This S1 dependence is included in the sensitivity calculation.)  The final fiducial mass is \DSfDdFiducialMass, with most of the uncertainty coming from the uncertainty in the thermal contraction of the Teflon reflector.

The \FNine\ acceptance vs.~\SOne\ is determined from the \FNine\ parametrization as described in Sec.~\ref{sec:Calibration}.  Figure~\ref{fig:Acceptance} shows acceptance vs.~\SOne\ for the analysis cuts.

Having designed a box to achieve our background target using cuts with understood acceptance, we proceeded to unblinding.

\begin{figure}[!t]
\includegraphics[width=0.5\textwidth]{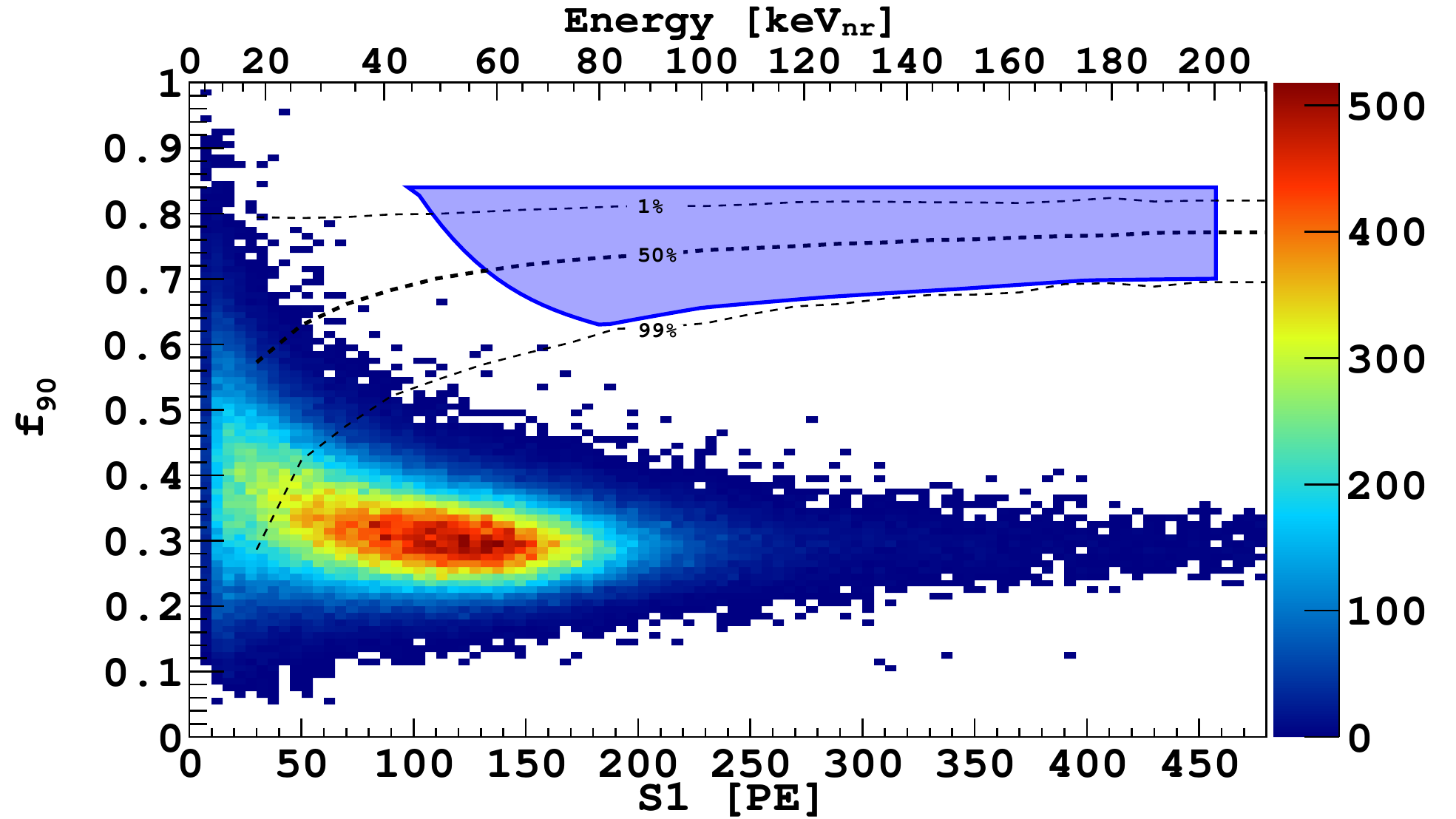} 
\caption{Observed events in the \FNine\ vs. \SOne\ plane surviving all cuts  in the energy region of interest.  The solid blue outline indicates the \DM\ search region.  The 1\%, 50\%, and 99\% \FNine\ acceptance contours for nuclear recoils, as derived from fits to our \AmBe\ calibration data, are shown as the dashed lines.}
\label{fig:DMSearchRegularCut}
\end{figure}

\section{Unblinding}
\label{sec:Unblinding}

Unblinding consisted of changing the access permissions of the open SLAD (see Sec.~\ref{sec:Blinding}), the blinded versions of which had been used for the background predictions, and running the analysis code applying all cuts to it.  Figure \ref{fig:DMSearchRegularCut} shows \FNine\ vs. \SOne\ after all analysis cuts.  With the analysis cuts applied and the data fully unblinded, no events are observed in the pre-defined \DM\ search region.  

After unblinding, we tabulated events surviving each cut, as shown in Table~\ref{tab:AccSummary}.  The order that the cuts were applied is not meaningful -- the order shown in the table was chosen to be informative.  Each of the last two events in Table~\ref{tab:AccSummary} was cut by both the prompt and delayed veto cuts.  They are the events in the box in Fig.~\ref{fig:DSfDdNeutrons} labeled ``Radiogenic" and ``Fission candidate". 

\section{WIMP Sensitivity and Limit}
\label{sec:Sensitivity}

\begin{figure}[!t]
\includegraphics[width=0.5\textwidth]{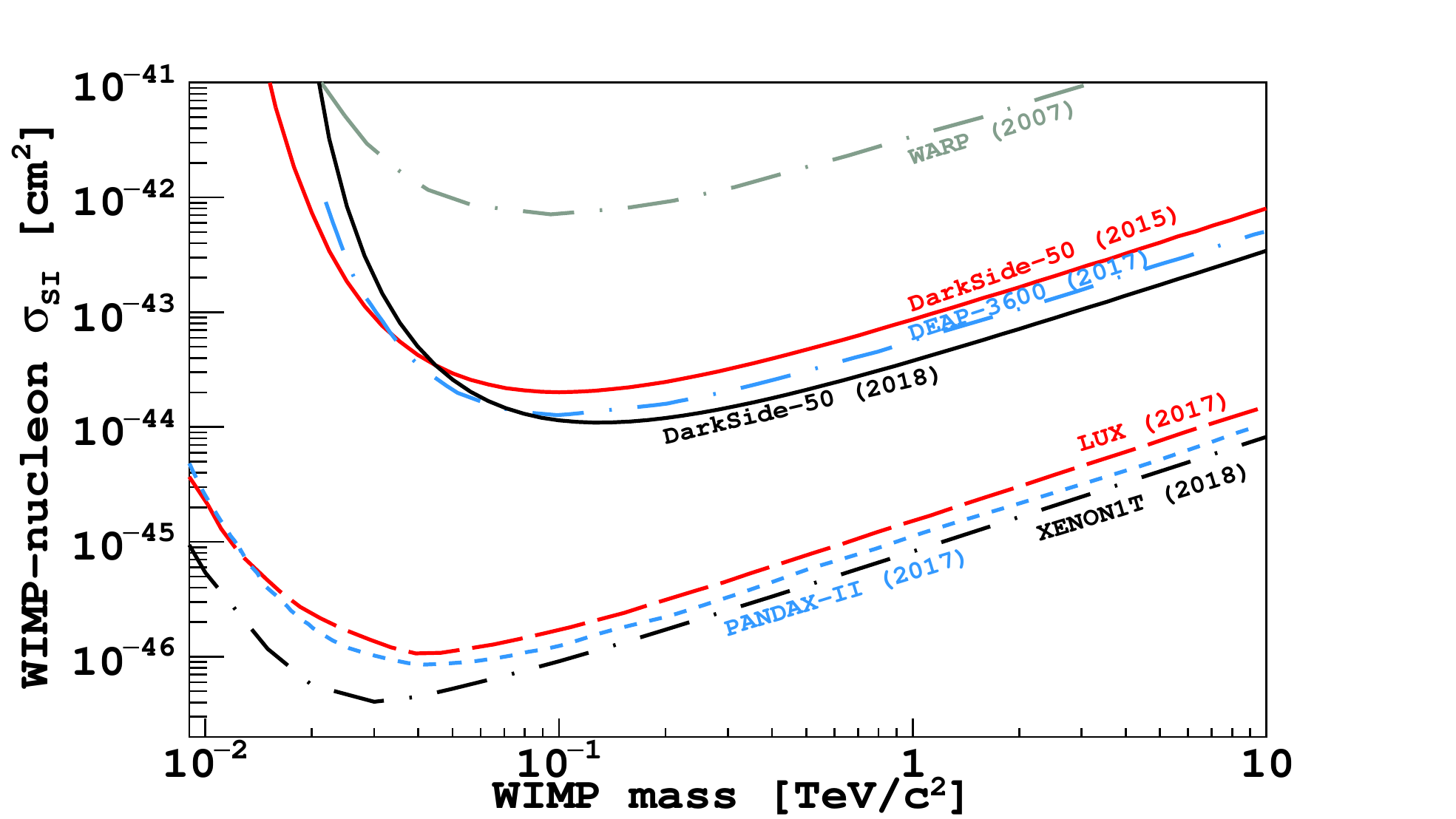}
\caption{Spin-independent \DM-nucleon cross section \SI{90}{\percent}~C.L. exclusion limits from the analysis detailed in this paper, compared to our previous result~\cite{Agnes:2016fz} and selected results from other experiments using argon (WARP~\cite{Benetti:2008kd}, DEAP-3600~\cite{Amaudruz:2018gr}) and xenon (LUX~\cite{Akerib:2017kg}, XENON1T~\cite{Aprile:2018vb}, PandaX-II~\cite{Cui:2017kg}).}
\label{fig:ExclusionLimits}
\end{figure}

A limit on spin-independent \DM-nucleon scattering is derived assuming the standard isothermal WIMP halo model, with $v_{\text{escape}} = \SI{544}{\km\per\sec}$~\cite{Smith:2007fi}, $v_{0} = \SI{220}{\km\per\sec}$~\cite{Smith:2007fi}, $v_{\text{Earth}} = \SI{232}{\km\per\sec}$~\cite{Savage:2009hi}, and $\rho_{\text{DM}} = \SI{0.3}{\GeV\per\square\c\per\cubic\cm}$~\cite{Savage:2006bt}.  The background- and signal-free result is consistent with up to \num{2.3} \DM-induced scatters (\SI{90}{\percent} C.L.), which sets an upper limit on the spin-independent DM-nucleon cross-section at \DSfDdHundredGeVLim\ (\DSfDdOneTeVLim, \DSfDdTenTeVLim) for \SI{100}{\GeV\per\square\c} (\SI{1}{\TeV\per\square\c}, \SI{10}{\TeV\per\square\c}) \DM\ particles.  The minimum upper limit is \DSfDdMinLim\ at  \DSfDdMinLimMass.  Figure~\ref{fig:ExclusionLimits} compares this limit to those obtained by other experiments.

\begin{figure}[!h]
\includegraphics[width=0.5\textwidth]{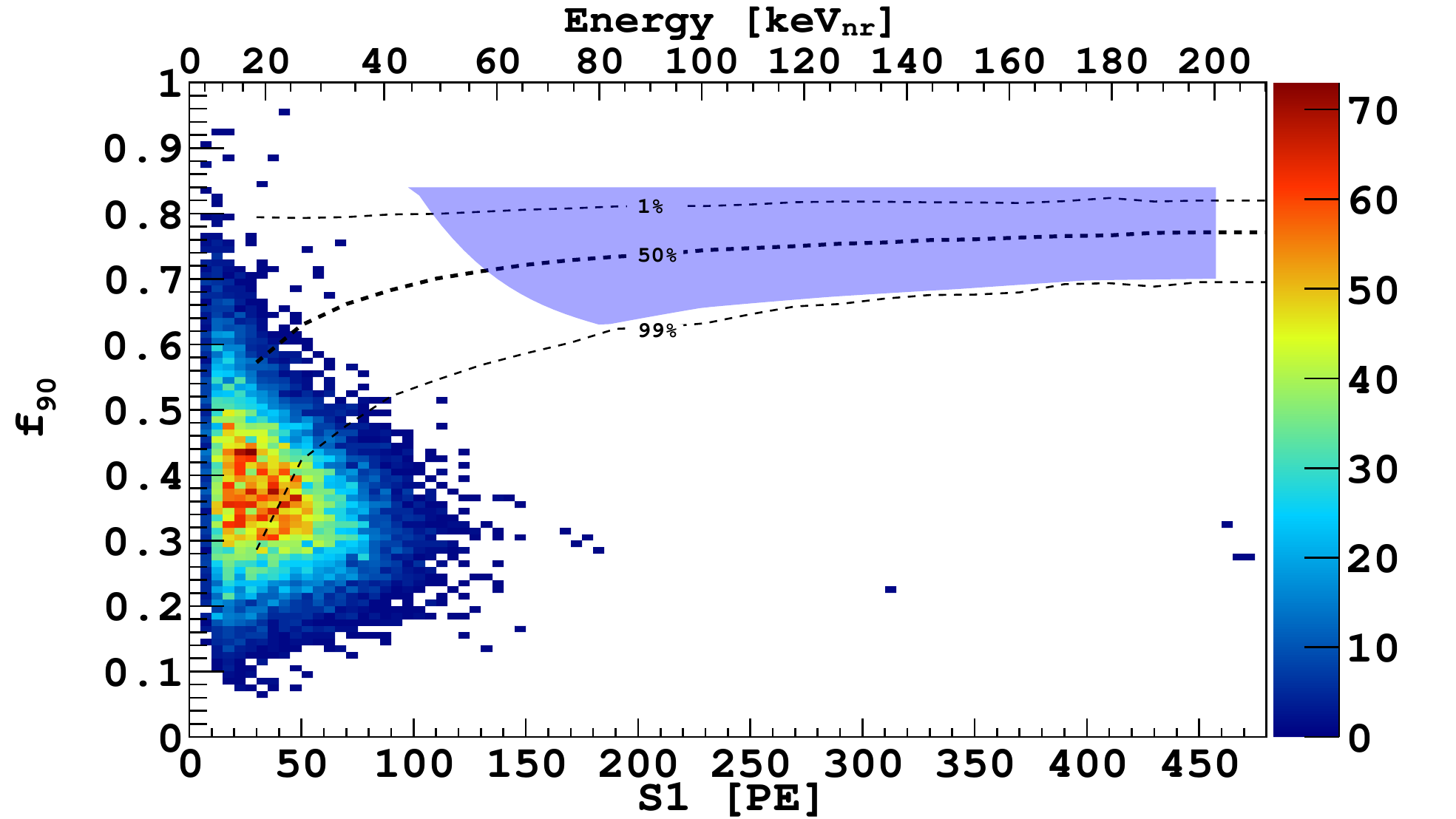} 
\caption{Distribution of events in the \FNine\ vs \SOne\ plane that survive all analysis cuts and that in addition survive tightened radial and \STwo/\SOne\ cuts (see text for details).}
\label{fig:DMSearchExtendedCut}
\end{figure}

Figure~\ref{fig:DMSearchExtendedCut} demonstrates available improvements in background rejection, which we do not use in this analysis.  If we require \STwo/\SOne\  lower than the median value for nuclear recoils and also radial fiducialization to about 8~cm from the wall ($r<10$~cm), we obtain an even greater separation between the events surviving the selection and the previously defined \DM\ search region.  In a multi-tonne detector~\cite{Aalseth:2018gq}, these cuts would provide exceptional background rejection at the cost of an affordable loss in detection efficiency.

\vspace{1cm}
\begin{acknowledgments}
The \DS\ Collaboration offers its profound gratitude to the \LNGS\ and its staff for their invaluable technical and logistical support.  We also thank the Fermilab Particle Physics, Scientific, and Core Computing Divisions.  Construction and operation of the \DSf\ detector was supported by the U.S. National Science Foundation (NSF) (Grants \grant{PHY}{0919363}, \grant{PHY}{1004072}, \grant{PHY}{1004054}, \grant{PHY}{1242585}, \grant{PHY}{1314483}, \grant{PHY}{1314501}, \grant{PHY}{1314507}, \grant{PHY}{1352795}, \grant{PHY}{1622415}, and associated collaborative grants \grant{PHY}{1211308} and \grant{PHY}{1455351}), the Italian Istituto Nazionale di Fisica Nucleare, the U.S. Department of Energy (Contracts \grant{DE}{FG02-91ER40671}, \grant{DE}{AC02-07CH11359}, and \grant{DE}{AC05-76RL01830}), the Russian Science Foundation (Grant \grant{16}{12-10369}), the Polish NCN (Grant \grant{UMO}{2014/15/B/ST2/02561}) and the Foundation for Polish Science (Grant \grant{Team2016}{2/17}).  We also acknowledge financial support from the French Institut National de Physique Nucl\'eaire et de Physique des Particules (IN2P3), from the UnivEarthS Labex program of Sorbonne Paris Cit\'e (Grants \grant{ANR}{10-LABX-0023} and \grant{ANR}{11-IDEX-0005-02}), and from the S\~ao Paulo Research Foundation (FAPESP) (Grant \grant{2016/09084}{0}).  Isotopes used in this research were supplied by the United States Department of Energy Office of Science by the Isotope Program in the Office of Nuclear Physics.
\end{acknowledgments}

\bibliography{./ds}
\bibliographystyle{./ds}
\end{document}